\documentclass[12pt,english,amsmath,amssymb,aps,pra,letterpaper,preprint,superscriptaddress]{revtex4-1}
\usepackage{graphicx}
\usepackage[T1]{fontenc}
\usepackage[latin9]{inputenc}
\usepackage[letterpaper]{geometry}
\usepackage{units}
\usepackage{esint}
\usepackage{braket}
\usepackage{numprint}

\makeatletter

\usepackage{mathrsfs}
\usepackage{bbm}

\makeatother

\usepackage{babel}

\begin{document}

\title{Population and Coherence Dynamics in Light Harvesting Complex II (LH2) }

\author{Shu-Hao Yeh}
\affiliation{Department of Chemistry and Birck Nanotechnology Center, Purdue University,
West Lafayette, IN 47907, USA}
\author{Jing Zhu}
\affiliation{Department of Chemistry and Birck Nanotechnology Center, Purdue University,
West Lafayette, IN 47907, USA}
\author{Sabre Kais}
\thanks{Corresponding author, kais@purdue.edu}
\affiliation{Department of Chemistry and Birck Nanotechnology Center, Purdue University,
West Lafayette, IN 47907, USA}
\affiliation{Qatar Environment and Energy Research Institute, Qatar Foundation,
Doha, Qatar}

\begin{abstract}
The electronic excitation population and coherence dynamics in the chromophores of the photosynthetic light harvesting complex 2 (LH2) B850 ring from purple bacteria (\textit{Rhodopseudomonas acidophila}) have been studied theoretically at both physiological and cryogenic temperatures. Similar to the well-studied Fenna-Matthews-Olson (FMO) protein, oscillations of the excitation population and coherence in the site basis are observed in LH2 by using a scaled hierarchical equation of motion (HEOM) approach. However, this oscillation time (300 fs) is much shorter compared to the FMO protein (650 fs) at cryogenic temperature. Both environment and high temperature are found to enhance the propagation speed of the exciton wave packet yet they shorten the coherence time and suppress the oscillation amplitude of coherence and the population. Our calculations show that a long-lived coherence between chromophore electronic excited states can exist in such a noisy biological environment.
\end{abstract}
\maketitle

\section{Introduction}

Photosynthesis is the single most crucial biochemical process in plants and algae since it transforms electromagnetic energy into chemical energy. Due to its importance in converting solar energy to a biologically utilizable form \cite{ProbePhotosyn2000}, the molecular mechanism of photosynthesis has been widely studied, especially in photosynthetic bacteria such as green sulphur bacteria, purple phototrophic bacteria, green filamentous bacteria, cyanobacteria, and heliobacteria \cite{Chlorophylls, AnoxyBacBook, MolMechPhoto}.

The light harvesting process consists of several steps. First, the incoming photon excites the electronic state of pigments in the antenna complex of the photosystem. Then the excited state energy will be transported to the reaction center of the photosystem. When enough excitation energy has been collected, a charge separation will be triggered by one or a pair of bacteriochlorophylls (BChls) at the reaction center; this will further induce an electron transfer via an electron transport chain and generate a proton gradient across the membrane. This proton gradient drives ATP synthase, which produces the most common energy form (ATP) utilized in organisms. Several reviews of the light harvesting process in purple bacteria are provided here \cite{Sundstrom1999, Yang2001, Grondelle2006, Cogdell2006, Hu2002}.

In purple bacteria - a model organism in photosynthesis studies - the photosynthetic units are located on the inner membrane and are comprised of two types of light harvesting membrane pigment-protein complexes. One is light harvesting complex 2 (LH2), which is the peripheral antenna that mainly absorbs light. The other is light harvesting complex 1 (LH1), which surrounds a reaction center and forms the LH1-RC complex which acts as the core antenna. The main pigments that absorb light and contribute to the excitation energy transfer in these antenna complexes are carotenoid (Car) and bacteriochlorophyll (BChl).

One of the most intriguing features of photosynthesis is the near-unity efficiency of energy conversion from light to charge \cite{MolMechPhoto}. The detailed mechanism of this delicate biological process remains elusive. The most well-studied systems are the Fenna-Matthews-Olson (FMO) complex of green sulfur bacteria and the LH2 complex of purple bacteria. The FMO complex transfers the exciton energy from the light harvesting complex to the reaction center (RC), thus plays the role of an intermediate candidate which passes the exciton energy with an efficiency of almost 100\% \cite{Andrew-book}. A long-lived quantum coherence was experimentally found in the pigment excited electronic states \cite{Engel-2007}, implying that quantum mechanical effects may be involved in the excitation energy transfer. Recently, Aspuru-Guzik et al. studied the effects on efficiency of the combination of quantum coherence and environment interaction via a non-Markovian approach developed from the Lindblad form \cite{Alan-2008, Alan-2009, Alan-JPC-2009, Alan2011Com}. Meanwhile, Ishizaki and coworkers \cite{Fleming-JCP-2009, Fleming-PNAS-2009} reproduced the experimentally detected population oscillation time observed by Engel and coworkers \cite{Engel-2007} via the hierarchical equation of motion (HEOM) approach. The excitation energy transfer pathway in the FMO complex has been investigated by Skochdopole et al \cite{Mazziotti2011}. The role of pairwise entanglement in FMO has also been investigated by Sarovar et al \cite{Sarovar:2010p6945}. Recently we applied the scaled HEOM method developed by Shi et al. \cite{Shi2009a} to the study of the quantum evolution of the FMO complex, showing that the scaled HEOM method is computationally more stable and accurate \cite{Ourpaper}. In addition, we also investigated the role of multipartite entanglement by direct computation of the convex roof optimization \cite{Ourpaper2}. In addition, there are numerous theoretical studies on modelling the dynamics of excitation energy transfer in the FMO system \cite{kramerGPU,Reichman2011,PlenioDMRG2,coker2010, coker2011,Mazziotti2011a,Thorwart2011,Lloyd2011,Lloyd2011a,Lloyd2011b,cao2010,cao2011,Silbey2010,Silbey2011,Ratner2011,Mukamel2010a}.

For LH2 in purple bacteria, a long-time quantum coherence in the excited states of bacteriocholophylls (BChl) has also been found experimentally \cite{Scholes-2010, Sundstrom1999, Chachisvilis1997, Harel2012}, and the excitation energy transfer dynamics have been investigated theoretically \cite{Strumpfer2009a, Strumpfer2011}. In this article we employed the scaled HEOM approach to study the quantum evolution in the LH2 complex, and a quantitative description of the long-time quantum coherence is provided. The paper is organized as follows. In the method section, the theoretical framework of scaled HEOM as well as the coherence will be introduced. Next, the simulation result and the physical explanation for both population and coherence evolution will be investigated.

\newpage

\section{Method}

The system studied here is the BChls in the LH2 B850 ring of purple bacteria \textit{Rhodopseudomonas} (\textit{Rps.}) \textit{acidophila}  \cite{Papiz2003} (PDBID: 1NKZ). We employed the open quantum system method to explore the electronic excitation dynamics of this complex. The 18 B850 BChls of the LH2 ring are treated as the system, which couples to the environment modelled by an infinite set of harmonic oscillators. They interact with each other via system-environment coupling, and the total Hamiltonian can be described as:
\begin{equation}
\mathcal{H}=\mathcal{H_{S}}+\mathcal{H}_{B}+\mathcal{H}_{SB}, \label{eq:Htot}
\end{equation}
where $\mathcal{H_{S}}$, $\mathcal{H}_{B}$ and $\mathcal{H}_{SB}$ represent the Hamiltonian of the system, environment and system-environment coupling respectively.

\subsection{The Structure of the LH2 Complex}

The molecular structure of LH2 is highly symmetric, and usually possesses $C_8$ or $C_9$ symmetry depending on the different species of bacteria. For example, in \textit{Rhodospirillum} (\textit{Rs.}) \textit{molischianum} \cite{Koepke1996} and \textit{Rps. acidophila} \cite{MCDERMOTT1995, Papiz2003} LH2 has an eight-fold and nine-fold symmetry respectively. In \textit{Rps. acidophila} each subunit is constructed by inner $\alpha$ and outer $\beta$-peptides and containing 3 BChls and 1 Car. In the LH2 absorption spectrum, there are two significant peaks at 800 and 850 $\unit{nm}$ due to the $Q_Y$ excitation of BChls. For each subunit there is one BChl that belongs to the 800 nm absorption peak and two BChls that contribute to the 850 nm peak. Due to symmetry, both of them form ring structures which are called the B800 and B850 ring, respectively.

\begin{figure}
\centering
\includegraphics[scale=0.6]{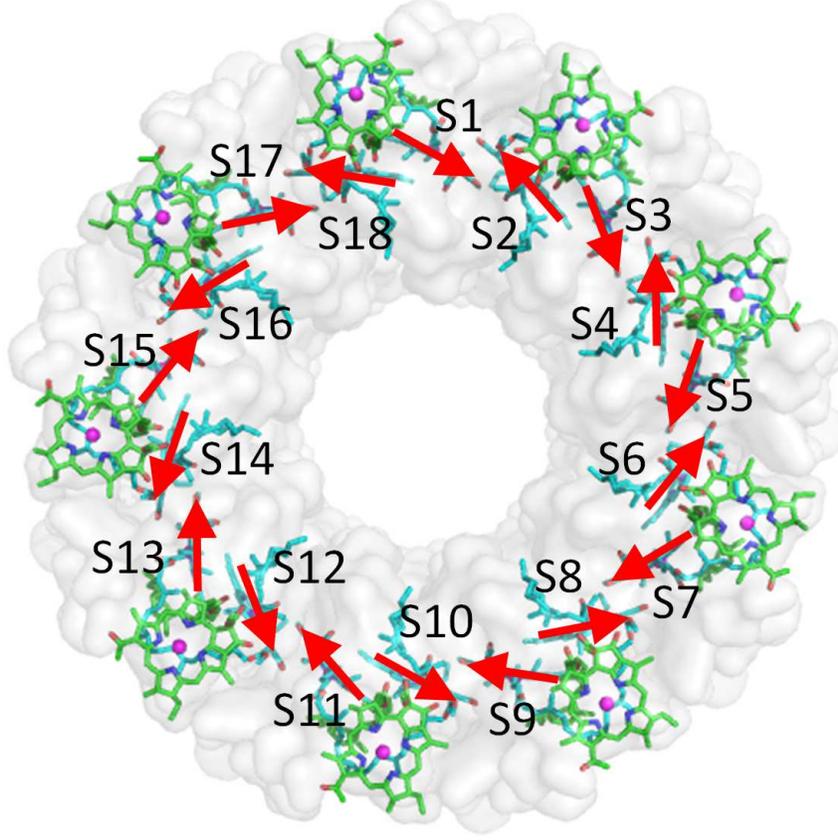}
\caption{\label{fig:LH2_structure} (Color) The top view (from cytoplasmic to periplasmic side) of 2.0~\AA\ resolution LH2 structure of \textit{Rhodopseudomonas acidophila}. The 9 B800 and 18 B850 BChls are shown in green and cyan respectively, and the purple spheres represent Magnesium of the BChls. The 18 BChls are labelled clockwise from S1 to S18, and the transition dipole of $Q_Y$ excitation for each BChl is represented by red arrows.}
\end{figure}

In this paper, we will only study the quantum evolution of the B850 ring. We plan to explore the role of both rings (B800 and B850) during the energy transfer in the future.
B850 ring is constituted by 18 BChls in nine transmembrane $\alpha$, $\beta$-polypeptide heterodimers and has a $C_9$ symmetry \cite{PurpleBac}. The arrangement of 18 B850 BChls is shown in Fig. \ref{fig:LH2_structure}.

All 18 B850 BChls are marked clockwise, and the distance between BChls is defined by the distance between the Magnesium atoms corresponding to each BChl. The distance between adjacent B850 BChls in the same and different dimers are 9.4~\AA\ and 9.1~\AA\ respectively \cite{Papiz2003}. The most popular theoretical models in this area can be organized into three distinct groups, the Redfield equation, F\"{o}rster theory and hierarchical equation of motion (HEOM).  The Redfield equation \cite{redfield1, redfield2} assumes that the coupling between the system and environment is weak compared with the coupling within the system. Mathematically, we can treat the environment effects as a perturbation upon the system. The assumptions for the F\"{o}rster theory are in total opposition to that of the Redfield equation \cite{ForsterBook1, ForsterBook2}. It assumes strong coupling between the system and environment so the interactions within the system can be neglected. As a result, the mathematical calculation can be simplified by treating the system Hamiltonian as a diagonal matrix. 
The HEOM does not require both assumptions. It can be applied to both strong and weak system-environment coupling, making it an accurate and comprehensive model \cite{Tanimura, Tanimura-Kubo-1989, Tanimura2, Shi2009a}. However, the numerical implementation requires a lot of computational resources. The electrostatic interaction between neighbouring B850 BChls is relatively strong due to the short distance between them, which makes the BChl electronic coupling at the same order as the electron-environment coupling. As a result, the HEOM is an ideal framework for this complex system. Our group has successfully employed the scaled HEOM approach to study the population beating and multiple particles entanglement evolution in the FMO complex \cite{Ourpaper, Ourpaper2}. The simulation results also indicate that the scaled HEOM is helpful in reducing the required computational resources without losing any resolution or physical meaning.

\subsection{Hamiltonian of the system}

Under a single exciton basis set, the system Hamiltonian can be written as \cite{Renger2001}
\begin{equation}
\mathcal{H}_{S}=\sum_{j=1}^{N}\varepsilon_{j}\,\ket{j}\bra{j}+\sum_{j\neq k}J_{jk}\left(\ket{j}\bra{k}+\ket{k}\bra{j}\right), \label{eq:HS}
\end{equation}
where $\varepsilon_{j}$ represents the $Q_Y$ excitation energy of the $j$th BChl and $J_{jk}$ denotes the excitonic coupling between BChl $j$ and $k$. There are only two different excitation energy for these 18 BChls. One is \numprint{12457.66} cm$^{-1}$ and the other is \numprint{12653.66} cm$^{-1}$ \cite{PurpleBac}. They arrange alternatively $9$ times \cite{Hu1997}. In our simulation, we treated the energy of BChl 1 (S1) as \numprint{12457.66} cm$^{-1}$ and thus BChl 2 (S2) equals \numprint{12653.66} cm$^{-1}$.

The non-nearest neighbor coupling between $j$ and $k$ was treated by dipole-dipole interaction, expressed as

\begin{equation}
J_{jk}=\frac{|p|^2}{4\pi \epsilon_0}
\frac{\hat{p}_j\cdot\hat{p}_k - 3\left(\hat{p}_j\cdot\hat{r}_{jk}\right)\left(\hat{p}_k\cdot\hat{r}_{jk}\right)}{\left|\vec{r}_{jk}\right|^{3}}
, \label{VijO}
\end{equation}

where $\hat{p}_j$ and $\hat{p}_k$ represents the unit vector of the $j$-th and $k$-th BChl dipole moment. $|\vec{r}_{jk}|$ indicates the distance between the Magnesium of the two BChls and $\hat{r}_{jk}$ is the unit vector along $\vec{r}_{jk}$. The same dipole moment has been used for all BChls with the value of $|p|^2 / 4\pi\epsilon_0$ set to \numprint{348000} \AA$\unit{^{3} cm^{-1}}$ \cite{Sener2007}. For the two adjacent BChls in the same (S1 and S2) and different dimers (S1 and S18), the couplings are set as $363\;\unit{cm^{-1}}$ and $320\;\unit{cm^{-1}}$ respectively \cite{Sener2007, Tretiak2000}.

\subsection{The System-Environment Coupling}
The environment is described as a phonon bath, modelled by an infinite set of harmonic oscillators:
\begin{equation}
\mathcal{H}_{B} =\sum_{j=1}^{N}\mathcal{H}_{B}^{j}=\sum_{j=1}^{N}\sum_{\xi=1}^{N_{jB}}\frac{P_{j\xi}^{2}}{2m_{j\xi}}+\frac{1}{2}m_{j\xi}\omega_{j\xi}^{2}x_{j\xi}^{2}\;, \label{eq:HB}
\end{equation}
where $m_{j\xi}$, $\omega_{j\xi}$, $P_{j\xi}$, $x_{j\xi}$ are mass, frequency, momentum and the position operator of the $\xi$-th harmonic bath associate with the $j$-th BChl respectively.

The Hamiltonian of the environment $\left( \mathcal{H}_{B} \right)$ and system-environment coupling $\left( \mathcal{H}_{SB} \right)$ can be written as:

\begin{equation}
\mathcal{H}_{SB} =\sum_{j=1}^{N}\sum_{\xi}c_{j\xi}\ket{j}\bra{j}x_{j\xi} = \sum_{j=1}^{N}\mathcal{V}_{j}F_{j}, \label{eq:HSB}\\
\end{equation}

where $\mathcal{V}_{j}=\ket{j}\bra{j}$ and $F_{j}=\sum_{\xi}c_{j\xi} x_{j\xi}$. $c_{j\xi}$ represents the system-bath coupling constant between the $j$-th BChl and $\xi$-th phonon mode. Here we assume each BChl is coupled to the environment independently.

The initial condition can be written as:
\begin{equation}
\chi(0)=\rho(0)\otimes\rho^{eq}_{B}, \label{eq:inistate}\\
\end{equation}
where $\chi$, $\rho$, and $\rho^{eq}_{B}$ are the total, system, and the bath density matrices, respectively.
The system is in its ground state $\ket{0}\bra{0}$ before being excited, and the environment is in a thermal equilibrium state. Once the system was excited by a laser pulse, the system suddenly changed to the specified state $\rho(0)$, while the environment still resides at equilibrium. After $t\,=\,0$, the system starts to evolve under the effect of the system-environment bath \cite{xu1,xu2}.
The time evolution of the system can be obtained by tracing the environment
$\rho\left(t\right)=\textrm{Tr}_{B}\left[\chi\left(t\right)\right]=
\textrm{Tr}_{B}\left[e^{\nicefrac{-i\mathcal{H}t}{\hbar}}\,\chi\left(0\right)\,e^{\nicefrac{i\mathcal{H}t}{\hbar}}\right].$

The correlation function under phonon bath can be written as:
\begin{align}
C_{j}\left(t\right) & =\frac{1}{\pi}\intop_{-\infty}^{\infty}d\omega D_{j}\left(\omega\right)\frac{e^{-i\omega t}}{1-e^{-\beta\hbar\omega}}\;,\label{eq:correlation}\\
D_{j}\left(\omega\right) & =\sum_{\xi}\frac{ c_{j\xi}^{2}\hbar}{2m_{j\xi}\omega_{j\xi}}\delta\left(\omega-\omega_{j\xi}\right)\;,\label{eq:spec density}
\end{align}
where $\beta=1/k_{B}T$. We replaced the original spectral
density (Eq. \ref{eq:spec density}) with the Drude spectral density
\begin{align}
D_{j}\left(\omega\right)=\frac{2\lambda\gamma}{\hbar}\frac{\omega}{\omega^{2}+\gamma^{2}}\;,\label{eq:Drude}
\end{align}
where $\lambda$ is the reorganization energy and $\gamma$ is the
Drude decay constant. The correlation function can be written as:
\begin{align}
C_{j}\left(t>0\right) & =\sum_{k=0}^{\infty}c_{k}\cdot e^{-v_{k}t}\;,\label{eq:correlationfunc}
\end{align}
 with $v_{0}=\gamma$, which is the Drude decay constant, $v_{k}=2k\pi/\beta\hbar$
when $k\geqslant1$ and $v_{k}$ is known as the Matsuraba frequency.
The constants $c_{k}$ are given by:
\begin{align}
c_{0}=\frac{\eta\gamma}{2}\left[\cot\left(\frac{\beta\hbar\gamma}{2}\right)-i\right]\;, \label{eq:c0}\\
c_{k}=\frac{2\eta\gamma}{\beta\hbar}\cdot\frac{v_{k}}{v_{k}^{2}-\gamma^{2}}\;\; \mathrm{for} \, k\geqslant1 \;. \label{eq:ck}
\end{align}

Using the scaled HEOM approach developed by Shi et al \cite{Shi2009a} as well as the Ishizaki-Tanimura truncation scheme \cite{Tanimura,Tanimura2}, the time evolution of the density operator becomes \cite{Ourpaper}:

\begin{multline}
\frac{d}{dt}\rho_{\boldsymbol{n}}=-\frac{i}{\hbar}\left[\mathcal{H}_{S},\;\rho_{\boldsymbol{n}}\right]-\sum_{j=1}^{N}\sum_{k=0}^{K}n_{jk}v_{k}\cdot\rho_{\boldsymbol{n}}-i\sum_{j=1}^{N}\sqrt{\left(n_{jk}+1\right)\left|c_{k}\right|}\,\left[\mathcal{V}_{j},\;\sum_{k}\rho_{\boldsymbol{n_{jk}^{+}}}\right]\\
-\sum_{j=1}^{N}\left(\sum_{m=K+1}^{\infty}\frac{c_{jm}}{v_{jm}}\right)\cdot\left[\mathcal{V}_{j},\,\left[\mathcal{V}_{j},\,\rho_{\boldsymbol{n}}\right]\right]-i\sum_{j=1}^{N}\sum_{k=0}^{K}\sqrt{\frac{n_{jk}}{\left|c_{k}\right|}}\;\left(c_{k}\mathcal{V}_{j}\,\rho_{\boldsymbol{n_{jk}^{-}}}-c_{k}^{*}\rho_{\boldsymbol{n_{jk}^{-}}}\mathcal{V}_{j}\right)\;,\label{eq:HEOM}
\end{multline}
where $\boldsymbol{n}$ is defined as one set of nonnegative integers
$\boldsymbol{n}\equiv\{n_{1},n_{2},\cdots,n_{N}\}=\{\{n_{10},n_{11}\cdots,n_{1K}\},\cdots,\{n_{N0,}n_{N1}\cdots,n_{NK}\}\}$.
$K$ is the truncation level of the correlation function and $\boldsymbol{n_{jk}^{\pm}}$
refers to change the value of $n_{jk}$ to $n_{jk}\pm1$ in the global
index $\boldsymbol{n}$. The sum of $n_{jk}$ is defined as tier $\mathcal{N}_{c}$,
$\mathcal{N}_{c}=\sum_{j,k}n_{jk}$. This is another truncation in
our numerical simulation. The density operator with all indices equal
to $0$ is the system's reduced density operator (RDO) while all other
operators are auxiliary density operators (ADOs).

Apart from the population evolution, the coherence evolution is also evaluated here. The coherence between site $i$ and $j$ is defined as $\left|\rho_{ij}\right|$

\subsection{Simulation Details}
During the numerical simulation, we tried different truncation levels for both the correlation function (K) and tiers ($\mathcal{N}_{c}$) and checked the relative convergence. The highest truncation levels we can reach are $K=2$ and $\mathcal{N}_{c}=3$.

The reorganization energy ($\lambda$) and Drude decay constant ($\gamma$) are set at $200~\unit{cm^{-1}}$ and $100~\unit{fs^{-1}}$ respectively \cite{Freiberg2009, Zerlauskiene2008}. The simulation at these truncation levels require at least 19.6 GB of memory, and for each integration step (1 fs) it takes about 70 s on a 2.3 GHz cpu core.

Compared with the FMO complex \cite{Ourpaper}, these truncation levels are relatively low as the system of the LH2 B850 BChls ring is much larger (18 $\times$ 18) than that of FMO situation (7 $\times$ 7).
The time evolution of the LH2 system (Eq.\eqref{eq:HEOM}) was calculated by the 4th order Runge-Kutta method with a fixed time step of 1 fs. We calculated the excitation dynamics for up to 1.5 ps.


\newpage

\section{Results and discussion}


\subsection{Exciton Population Dynamics}
\begin{figure}[h]
\centering
\includegraphics[scale=0.8]{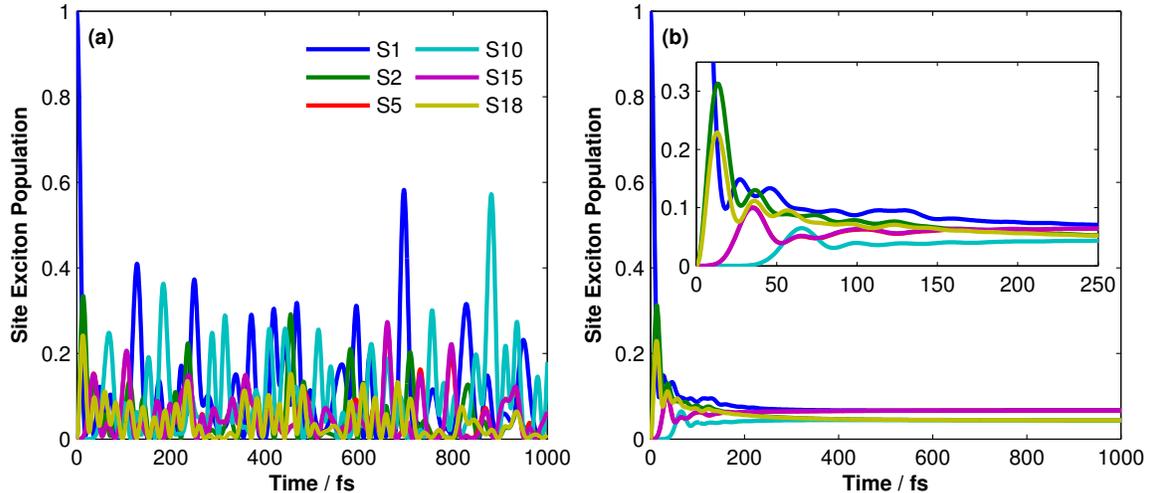}
\caption{\label{fig:300K_population} (Color) The exciton population dynamics of B850 bacteriochlorophylls (BChls) with site 1 (S1) initially excited. (a) The population evolution of S1, S2, S5, S10, S15, and S18 without dissipation (the system is isolated and uncoupled to bath). (b) The population dynamics of the same sites while the system is coupled to bath at room temperature T = 300 K. The coherent energy transfer lasts about 150 fs and the whole system is equilibrated after 400 fs. The inset is a magnification of the first 250 fs dynamics. }
\end{figure}

In the non-dissipative case (Fig. \ref{fig:300K_population}(a)), the population at S1 decreases rapidly and reaches its minimum value at $19\;\unit{fs}$, while the populations of S2 and S18 continually increase and arrive at their maximum value of 0.334, 0.243 at 14, 13 fs respectively. When $t\,=\,36\;\unit{fs}$, the populations of S5 and S15 become the maximally populated BChls. At 68 fs, the population of S10 dominates the whole LH2 system. Interestingly, the population of S1 hits another maximum value of 0.410 at 128 fs, which is almost twice of the time for the population passed from S1 to S10 (68 fs). This suggests that the excitation wave packet propagates both clockwise (towards S2) and counterclockwise (towards S18) and those two wave packets superpose at S10 and then continue traversing the entire ring.

For the dissipative case (Fig.~\ref{fig:300K_population}(b)), the populations change in a similar fashion to the non-dissipative situation for the first 50 fs. The simulation was run at temperature $T\,=\,300\;\unit{K}$. There is also population beating with the effect of the environment. The beating time of the population oscillation lasts roughly 150 fs. The populations of S2, S5 and S10 evolves to their first maximum value at 13, 35 and 66 fs respectively. Compared with the previous, non-dissipative case, these maxima are achieved simultaneously, which indicates the population oscillation of LH2 is due to the system itself rather than the environment. The environment plays a role in destroying the population beating, which is analogous to that in the FMO complex \cite{Ourpaper}. However, the population amplitude of each site and beating time are both smaller when compared to the FMO complex. This is reasonable since LH2 is much larger than the FMO complex.

\begin{figure}[h]
\centering
\includegraphics[scale=0.8]{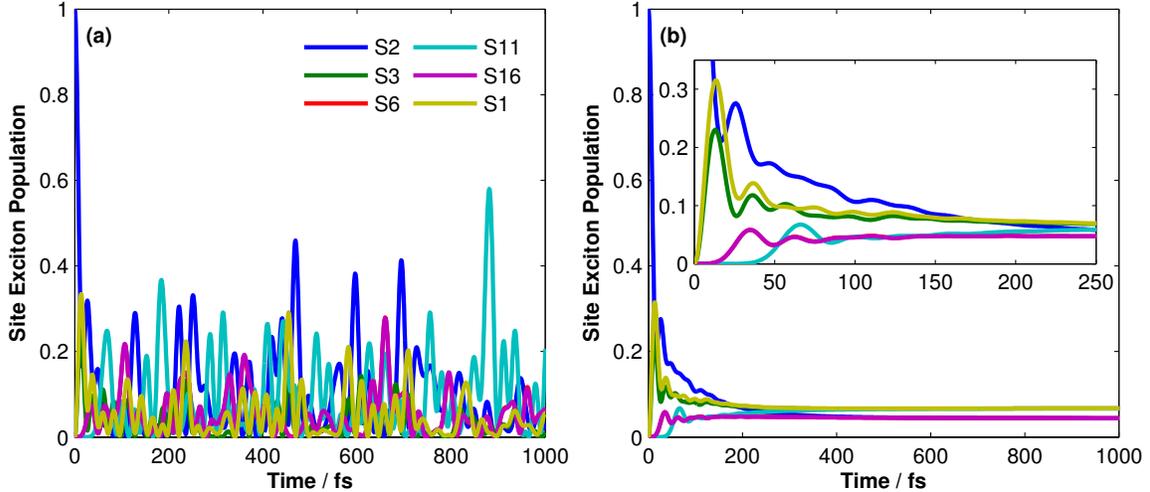}
\caption{\label{fig:300K_population_S2} (Color) The exciton population dynamics of B850 bacteriochlorophylls (BChls) with site 2 (S2) initially excited. (a) The population evolution of S1, S2, S3, S6, S11, and S16 without dissipation (the system is isolated and uncoupled to bath). (b) The population dynamics of the same sites while the system is coupled to bath at room temperature T = 300 K. The coherent energy transfer lasts about 150 fs and the whole system is equilibrated after 400 fs. The inset is a magnification of the first 250 fs-dynamics.}
\end{figure}

In contrast with the non-dissipative situation, the populations of all sites evolve to their equilibrium state during the time evolution. Furthermore, the difference for the peak arrival time also becomes larger as the evolution continues. It shows that the environment not only affects the population distribution but also facilitates the propagation speed of these exciton waves. The system arrives at the thermal equilibrium at 400 fs when $T\,=\,300\;\unit{K}$. Due to the symmetry of the system, all odd-number sites will finally equilibrate to the same population and similarly with even sites.
When two wave packets first reach S10, the local maximal population is about 1.6 times that of the first local maximum of the S5 and S15 populations, and the time variance of this peak increases as well. Although the system is highly symmetric, the coupling between sites is non-trivial and makes the oscillation pattern complex, which is quite different than in the FMO system. Using another initial condition (S2 fully excited), we found that in the non-dissipative case (Fig.~\ref{fig:300K_population_S2}(a)), the time when two opposite wave packets superimpose is identical to the situation of S1 initially excited.

For the dissipative case (Fig.~\ref{fig:300K_population_S2}(b)), the amplitude of the first population maximum of adjacent sites corresponds to the strength of coupling. Since the coupling between BChls in the same dimer is stronger than that between 2 BChls of different dimers, larger population can be found in the BChl in the same dimer. When S1 is initially excited, the population of S1 decreases rapidly to about 0.1 and oscillates around this value for about 150 fs. However, when taking S2 as the initial state, it drops quickly to 0.2 at first and then decreases toward its equilibrium by about 200 fs.
The population decrease of S2 is significantly larger than all other sites initially. However, the equilibrium population is found to be independent of initial conditions.

\subsection{Coherence Dynamics}

\begin{figure}[h]
\centering
\includegraphics[scale=0.8]{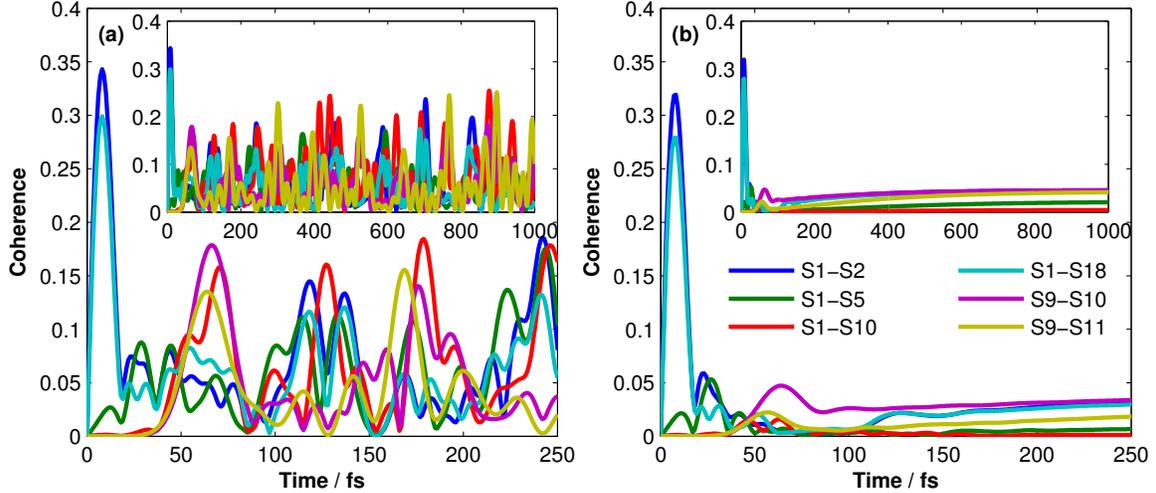}
\caption{\label{fig:300K_entanglement} (Color) The first 250 fs-coherence dynamics of B850 bacteriochlorophyll under 300 K with S1 initially excited (a) with or (b) without coupling to a phonon bath. The coherence decays as the distance between the sites increases. There is a significant long-time coherence between site 1 and site 2 excitons even the system is equilibrated. The inset shows the 1 ps-dynamics at the same condition.}
\end{figure}

From Fig.~\ref{fig:300K_entanglement}(a), it can be seen that the coherence lasted all the time for the non-dissipative case. The simulation result is valid as there are no exterior effects eliminating the coherence for an isolated system. Several important pairs are plotted in Fig.~\ref{fig:300K_entanglement}, which are S1-S2, S1-S5, S1-S10, S1-S18, S9-S10, and S9-S11. All the other pairs can be predicted through system symmetry. Although S1 and S10 are well separated in distance (about 52.75 {\AA}), their coherence is still significant.

\begin{figure}[h]
\centering
\includegraphics[scale=0.8]{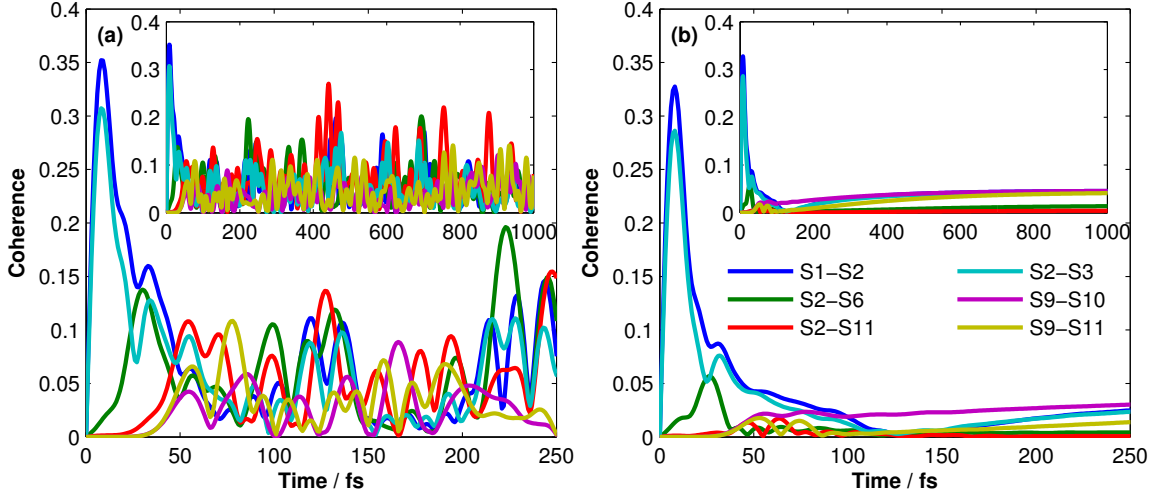}
\caption{\label{fig:300K_entanglement_S2} (Color) The coherence dynamics of B850 bacteriochlorophyll under 300 K with S2 initially excited (a) with or (b) without coupling to a phonon bath. The inset shows the 1 ps-dynamics at the same condition.}
\end{figure}

On the other hand, when a heat bath ($T\,=\,300\unit{K}$) is coupled to the system (Fig. ~\ref{fig:300K_entanglement}(b)), the coherence drops quickly during the evolution. For a quantitative view of decoherence, the second peak of S1-S5 coherence at 300 K and non-dissipative case has been compared, and we found that it suppresses 39.27\% at $t = 27$ fs when the system is dissipative. In addition, in the first peak of S9-S10 coherence (around $t = 64$ fs) the amplitude decreases 73.56\% and clearly demonstrates the decoherence effect. At the nearly-equilibrium state, the significant coherence (close to 0.1) occurs between adjacent sites (S1-S2, S1-S18, and S9-S10) and between second nearest sites (S9-S11). For the situation of S2 initially excited, the first peak of S2-S6 and the 1st peak of S9-S10 coherence show a decrease of 58.77\% (at $t\,=\,26$ fs)  and 48.63\% (at $t\,=\,56$ fs) respectively in the dissipative case (Fig.~\ref{fig:300K_entanglement_S2}(b)). This indicates that the environment plays an important role in reducing the coherence.

\begin{figure}[h]
\centering
\includegraphics[scale=0.8]{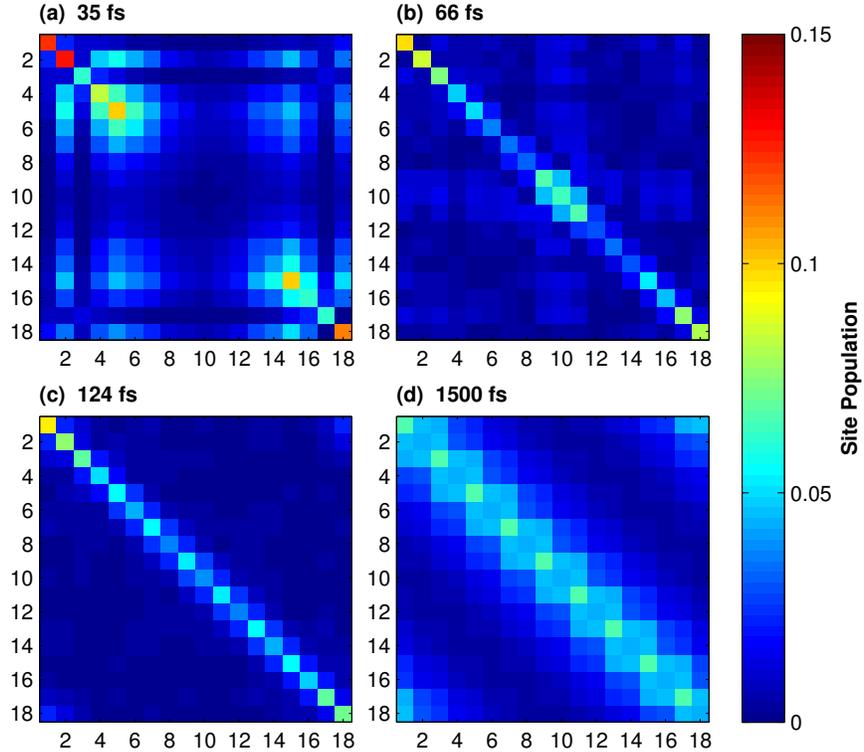}
\caption{\label{fig:300K_rho} (Color) The time evolution of density matrices (absolute value) of the 18-sites system at 300 K. The initial condition is S1 fully excited. (a) $t\,=\,35$ fs, when S5 and S15 have their maximum population. (b) $t\,=\,66$ fs, when S10 hits its maximum population. (c) $t\,=\,124$ fs, when both exciton wave packets traverse through the ring. (d) $t\,=\,1500$ fs, the system is in the equilibrium state.}
\end{figure}

To obtain a holistic picture of the information contained in the density matrix, the absolute value of all density matrix elements have been plotted (Fig.~\ref{fig:300K_rho}). The diagonal elements represent the exciton population, and the off-diagonal elements are the coherence between two sites. The initial state of Fig.~\ref{fig:300K_rho} is S1 fully excited. After $t\,=\,0$, the exciton wave packets propagate both clockwise (towards S2) and counterclockwise (towards S18) and merge at S10 at $t\,=\,66$ fs. Then, they traverse through the whole ring and back to S1 (at $\sim$124 fs). After that the system regains some coherence between adjacent sites. Even when the system is near its equilibrium (1.5 ps), there exists a significant coherence between several nearest neighbours (4 and 2 adjacent sites for odd and even number sites respectively) and thus provides additional evidence for the existence of long-lived coherence between adjacent chromophores in LH2 system.

%
\subsection{Temperature Effects on Exciton Dynamics}
In order to investigate the effects of temperature on exciton dynamics, the population and coherence dynamics have been studied at cryogenic temperature (77 K), this being the temperature at which most 2D electronic spectroscopy experiments have been performed.

\begin{figure}[h]
\centering
\includegraphics[scale=0.8]{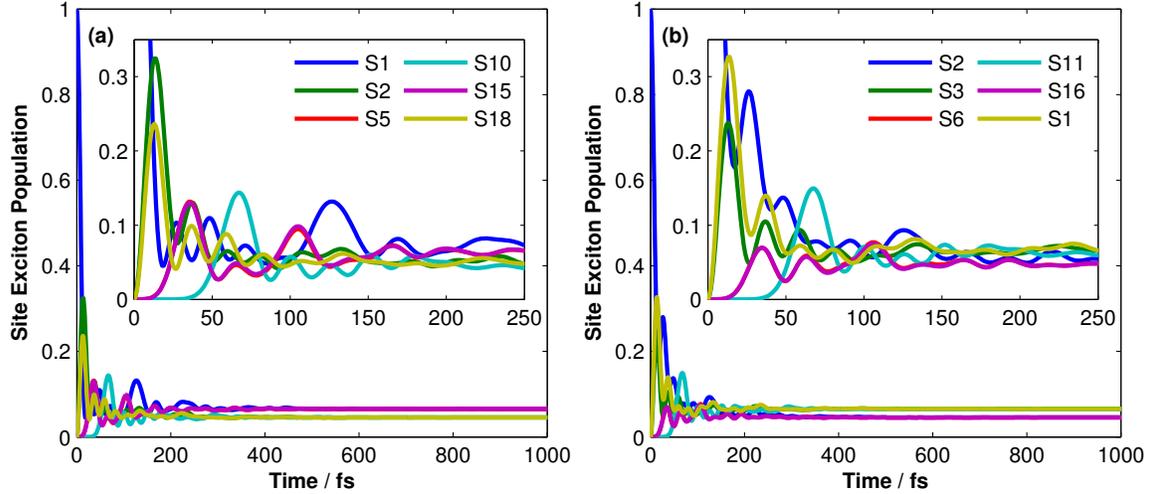}
\caption{\label{fig:77K_population} (Color) The excitation dynamics of B850 bacteriochlorophyll with (a) S1 or (b) S2 initially excited under 77 K. The coherent transfer between exciton lasts about 300 fs and the whole system is equilibrated after 600 fs. Longer population oscillation time is observed comparing to 300 K. Also around 127 fs (S1 initially excited) and 125 fs (S2 initially excited) the initial excited site has a local maximum with a larger width, which indicates when the two opposite exciton packets superpose. The inset is a magnification of the first 250 fs-dynamics}
\end{figure}

The exciton population dynamics at 77 K are shown in Fig. ~\ref{fig:77K_population}. Similar to the 300 K case, the exciton wave packets propagate both clockwise and counterclockwise. The time taken for two wave packets to merge at S10 is 67 fs, which is shorter than that of the isolated situation (68 fs), but longer compared with that for $T\,=\,300\;\unit{K}$ (66 fs). The time for the excitation wave packet returning to S1 is 127 fs, which is also located between that for isolated and $T\,=\,300K$ cases. The same phenomenon can be observed by checking the time when S5 and S10 have their 3rd maximum (non-dissipative: 106 fs; 77 K: 105 fs; 300 K: 104 fs). The environment increases the population propagation speed and higher temperature further improves the speed.



The population oscillation lasts over 300 fs at cryogenic temperature, which is almost twice that of at room temperature ($\sim$ 150 fs).
More quantitatively, one could also check the 1st maximum population ratio of S5 (or S15) to S10, which indicates the dissipative effect while the exciton wave packet travel from S5 to S10. At non-dissipative, 77 K and 300 K, these ratios are 1.634, 1.106, and 0.651 respectively. Since the first maximum population appears at almost the same time for all three conditions, it is valid to say that the dissipation rate gets larger with increasing temperature. A consequence of this is that the system will take longer to equilibrate ($\sim$600 fs at 77 K) at low temperature.
In other words, the system-environment coupling becomes stronger when the temperature increases; high temperature plays a positive role in driving the system into thermal equilibrium.

\begin{figure}[h]
\centering
\includegraphics[scale=0.8]{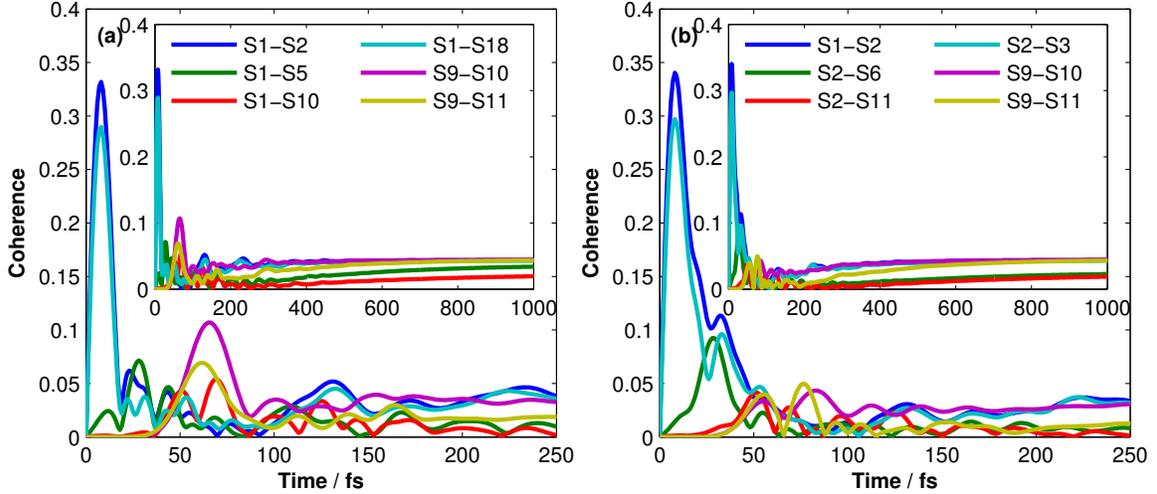}
\caption{\label{fig:77K_entanglement} (Color) The coherence dynamics of B850 bacteriochlorophyll under 77 K with (a) S1 or (b) S2 initially excited.  The oscillations of the coherence last about 250 fs, which is significantly longer than in the 300 K. The coherence between sites also decays as the distance increases. The inset shows the 1 ps-dynamics at the same condition.}
\end{figure}

The coherence dynamics at 77 K is shown in Fig.~\ref{fig:77K_entanglement}. The decay rate shows that the system decoheres slower compared to the case at 300 K. For the first 150 fs, all sites have relatively large coherence. On there other hand, the coherent energy transfer lasts about 300 fs, which is the same length as the population oscillation time. Again the second maximum of S1-S5 coherence (S1 initially excited) and the first maximum of S2-S6 coherence (S2 initially excited) have been compared to the non-dissipative case, and in both cases the coherence only be reduced by 18.64\% at $t = 28$ fs and 32.87\% at $t = 29$ fs, respectively. These results show that the dissipative effect occurs slower than the high temperature case. We conclude that the temperature plays a negative role in coherence conservation. Lower temperature is helpful for the system to maintain the coherence.

\begin{figure}[h]
\centering
\includegraphics[scale=0.8]{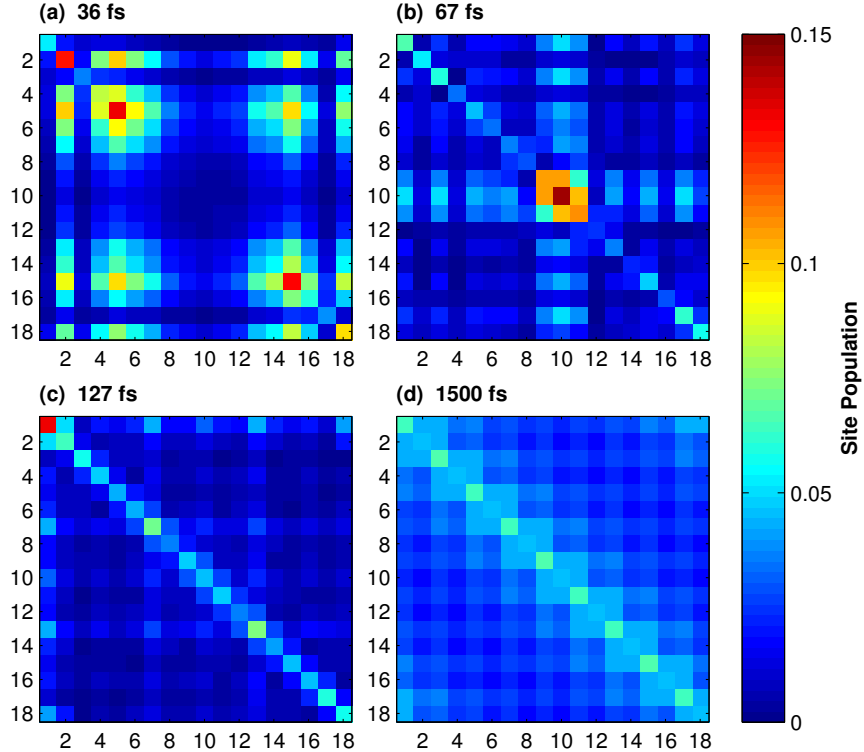}
\caption{\label{fig:77K_rho} (Color) The time evolution of density matrices (absolute value) of the 18-sites system at 77 K. The state with fully excited S1 is used as the initial condition. (a) $t\,=\,36$ fs, when S5 and S15 have their maximum population. (b) $t\,=\,67$ fs, when S10 has its maximum population. (c) $t\,=\,127$ fs, when both exciton wave packets traverse through about the whole ring. (d) $t\,=\,1500$ fs, the system is in a nearly equilibrated state.}
\end{figure}

After checking the absolute value of density matrix elements at 77 K (Fig.~\ref{fig:77K_rho}) starting from S1 excitation condition, we found that when S5 and S15 have their maximum, the S5-S15 coherence is significantly larger than that at room temperature. The same also occurs when S10 hits its first maximum, significant coherence appears between S10 and odd number sites all over the ring. This phenomenon is more apparent at low temperature. When the two opposite exciton wave packets propagate back to S1, in spite of a larger S1 population at low temperature, the coherence between S1-S7 and S10-S13 is also significant. For the final equilibrium state (at 1.5 ps) at 77 K, while the even number sites have coherence mainly with the nearest two neighbours, the odd number sites have coherence with its four nearest neighbours and the other odd number sites. This effect is absent in the high temperature case.

%
%
\section{Conclusions}

In summary, we employed the scaled hierarchical equation of motion (HEOM), which has been proved to be computationally more efficient \cite{Ourpaper}, in studying the population and the coherence evolution of the LH2 B850 for both cryogenic ($T\,=\,77\;\unit{K}$) and physiological ($T\,=\,300\;\unit{K}$) temperatures. Moreover, the short-time dynamics and the thermal equilibration of excitation energy transfer within the ring are systematically investigated by visualizing the reduced density operator.

Similar to the situation for the FMO complex, there are both population and coherence oscillations during the time evolution in the LH2 system. In addition, the oscillation time for population and coherence is almost the same for the same system. High temperatures shorten the coherence time and suppress both oscillation amplitude of coherence and population. However, the beating time for LH2 can last 150 fs and 300 fs at physiological and cryogenic temperatures, respectively. This is much shorter compared with the FMO complex (coherence oscillation can last 400 and 650 fs at the same temperature). Besides this, the exciton packet transfer both clockwise and counterclockwise during the evolution for LH2 system. The period for exciton transferred is 128, 127 and 124 fs for the non-dissipative, cryogenic and physiological temperature situations; this indicates that the high temperature increases the propagation speed of the exciton wave packet. Moreover, although the coherence oscillation vanishes rapidly, the coherence between adjacent BChls revives and remains in the system for a long time, which is consistent with the result of 2D electronic spectroscopy from other experimental studies.  This phenomenon may give us some clues about coherence could be maintained in such a noisy environment and may shed a light on designing a quantum system which can preserve coherence at room temperature.

Recently, the optimized HEOM coupled with the Drude spectral density has been proposed \cite{xu3,xu4}, which improves the efficiency in numerical calculations. It will be our future task to apply this approach to the current model for simulating larger systems, such as the LH2 B800-B850.

\section{Acknowledgement}

We would like to thank the NSF Center for Quantum Information and Computation for Chemistry (QIQC), Award number CHE-1037992, for financial support. S.Y acknowledges Hung-Yun Lin and Dr. Ganesh Subbarayan from the department of mechanical engineering, Purdue University, for the technical support and the computation resources.


\begin{thebibliography}{0}%
\makeatletter
\providecommand \@ifxundefined [1]{%
 \@ifx{#1\undefined}
}%
\providecommand \@ifnum [1]{%
 \ifnum #1\expandafter \@firstoftwo
 \else \expandafter \@secondoftwo
 \fi
}%
\providecommand \@ifx [1]{%
 \ifx #1\expandafter \@firstoftwo
 \else \expandafter \@secondoftwo
 \fi
}%
\providecommand \natexlab [1]{#1}%
\providecommand \enquote  [1]{``#1''}%
\providecommand \bibnamefont  [1]{#1}%
\providecommand \bibfnamefont [1]{#1}%
\providecommand \citenamefont [1]{#1}%
\providecommand \href@noop [0]{\@secondoftwo}%
\providecommand \href [0]{\begingroup \@sanitize@url \@href}%
\providecommand \@href[1]{\@@startlink{#1}\@@href}%
\providecommand \@@href[1]{\endgroup#1\@@endlink}%
\providecommand \@sanitize@url [0]{\catcode `\\12\catcode `\$12\catcode
  `\&12\catcode `\#12\catcode `\^12\catcode `\_12\catcode `\%12\relax}%
\providecommand \@@startlink[1]{}%
\providecommand \@@endlink[0]{}%
\providecommand \url  [0]{\begingroup\@sanitize@url \@url }%
\providecommand \@url [1]{\endgroup\@href {#1}{\urlprefix }}%
\providecommand \urlprefix  [0]{URL }%
\providecommand \Eprint [0]{\href }%
\providecommand \doibase [0]{http://dx.doi.org/}%
\providecommand \selectlanguage [0]{\@gobble}%
\providecommand \bibinfo  [0]{\@secondoftwo}%
\providecommand \bibfield  [0]{\@secondoftwo}%
\providecommand \translation [1]{[#1]}%
\providecommand \BibitemOpen [0]{}%
\providecommand \bibitemStop [0]{}%
\providecommand \bibitemNoStop [0]{.\EOS\space}%
\providecommand \EOS [0]{\spacefactor3000\relax}%
\providecommand \BibitemShut  [1]{\csname bibitem#1\endcsname}%
\let\auto@bib@innerbib\@empty
\end{thebibliography}%


\begin{thebibliography}{64}%
\makeatletter
\providecommand \@ifxundefined [1]{%
 \@ifx{#1\undefined}
}%
\providecommand \@ifnum [1]{%
 \ifnum #1\expandafter \@firstoftwo
 \else \expandafter \@secondoftwo
 \fi
}%
\providecommand \@ifx [1]{%
 \ifx #1\expandafter \@firstoftwo
 \else \expandafter \@secondoftwo
 \fi
}%
\providecommand \natexlab [1]{#1}%
\providecommand \enquote  [1]{``#1''}%
\providecommand \bibnamefont  [1]{#1}%
\providecommand \bibfnamefont [1]{#1}%
\providecommand \citenamefont [1]{#1}%
\providecommand \href@noop [0]{\@secondoftwo}%
\providecommand \href [0]{\begingroup \@sanitize@url \@href}%
\providecommand \@href[1]{\@@startlink{#1}\@@href}%
\providecommand \@@href[1]{\endgroup#1\@@endlink}%
\providecommand \@sanitize@url [0]{\catcode `\\12\catcode `\$12\catcode
  `\&12\catcode `\#12\catcode `\^12\catcode `\_12\catcode `\%12\relax}%
\providecommand \@@startlink[1]{}%
\providecommand \@@endlink[0]{}%
\providecommand \url  [0]{\begingroup\@sanitize@url \@url }%
\providecommand \@url [1]{\endgroup\@href {#1}{\urlprefix }}%
\providecommand \urlprefix  [0]{URL }%
\providecommand \Eprint [0]{\href }%
\providecommand \doibase [0]{http://dx.doi.org/}%
\providecommand \selectlanguage [0]{\@gobble}%
\providecommand \bibinfo  [0]{\@secondoftwo}%
\providecommand \bibfield  [0]{\@secondoftwo}%
\providecommand \translation [1]{[#1]}%
\providecommand \BibitemOpen [0]{}%
\providecommand \bibitemStop [0]{}%
\providecommand \bibitemNoStop [0]{.\EOS\space}%
\providecommand \EOS [0]{\spacefactor3000\relax}%
\providecommand \BibitemShut  [1]{\csname bibitem#1\endcsname}%
\let\auto@bib@innerbib\@empty
\bibitem [{\citenamefont {Mohammad~Yunus}\ and\ \citenamefont
  {Mohanty}(2000)}]{ProbePhotosyn2000}%
  \BibitemOpen
  \bibinfo {editor} {\bibfnamefont {U.~P.}\ \bibnamefont {Mohammad~Yunus}}\
  and\ \bibinfo {editor} {\bibfnamefont {P.}~\bibnamefont {Mohanty}},\ eds.,\
  \href@noop {} {\emph {\bibinfo {title} {Probing Photosynthesis: Mechanisms,
  Regulation, and Adaptation}}}\ (\bibinfo  {publisher} {Taylor and Francis,
  New York},\ \bibinfo {year} {2000})\ pp.\ \bibinfo {pages}
  {9--39}\BibitemShut {NoStop}%
\bibitem [{\citenamefont {Hawthornthwaite}\ and\ \citenamefont
  {Cogdell}(1991)}]{Chlorophylls}%
  \BibitemOpen
  \bibfield  {author} {\bibinfo {author} {\bibfnamefont {A.~M.}\ \bibnamefont
  {Hawthornthwaite}}\ and\ \bibinfo {author} {\bibfnamefont {R.~J.}\
  \bibnamefont {Cogdell}},\ }\href@noop {} {\emph {\bibinfo {title}
  {Chlorophylls}}},\ edited by\ \bibinfo {editor} {\bibfnamefont
  {H.}~\bibnamefont {Scheer}}\ (\bibinfo  {publisher} {CRC Press, Boca Raton},\
  \bibinfo {year} {1991})\ pp.\ \bibinfo {pages} {493--528}\BibitemShut
  {NoStop}%
\bibitem [{\citenamefont {Zuber}\ and\ \citenamefont
  {Cogdell}(1995)}]{AnoxyBacBook}%
  \BibitemOpen
  \bibfield  {author} {\bibinfo {author} {\bibfnamefont {H.}~\bibnamefont
  {Zuber}}\ and\ \bibinfo {author} {\bibfnamefont {R.}~\bibnamefont
  {Cogdell}},\ }\href@noop {} {\emph {\bibinfo {title} {Anoxygenic
  Photosynthetic Bacteria}}},\ edited by\ \bibinfo {editor} {\bibfnamefont
  {M.~M.}\ \bibnamefont {R.E.~Blankenship}}\ and\ \bibinfo {editor}
  {\bibfnamefont {C.}~\bibnamefont {Bauer}}\ (\bibinfo  {publisher} {Kluwer
  Academic, Dordrecht},\ \bibinfo {year} {1995})\ pp.\ \bibinfo {pages}
  {315--348}\BibitemShut {NoStop}%
\bibitem [{\citenamefont {Blankenship}(2002)}]{MolMechPhoto}%
  \BibitemOpen
  \bibinfo {editor} {\bibfnamefont {R.~E.}\ \bibnamefont {Blankenship}},\ ed.,\
  \href@noop {} {\emph {\bibinfo {title} {Molecular Mechanisms of
  Photonsynthesis}}}\ (\bibinfo  {publisher} {London: Blackwell Sci},\ \bibinfo
  {year} {2002})\BibitemShut {NoStop}%
\bibitem [{\citenamefont {Sundstrom}, \citenamefont {Pullerits},\ and\
  \citenamefont {van Grondelle}(1999)}]{Sundstrom1999}%
  \BibitemOpen
  \bibfield  {author} {\bibinfo {author} {\bibfnamefont {V.}~\bibnamefont
  {Sundstrom}}, \bibinfo {author} {\bibfnamefont {T.}~\bibnamefont
  {Pullerits}}, \ and\ \bibinfo {author} {\bibfnamefont {R.}~\bibnamefont {van
  Grondelle}},\ }\href {\doibase 10.1021/jp983722+} {\bibfield  {journal}
  {\bibinfo  {journal} {J Phys Chem B}\ }\textbf {\bibinfo {volume} {103}},\
  \bibinfo {pages} {2327} (\bibinfo {year} {1999})}\BibitemShut {NoStop}%
\bibitem [{\citenamefont {Yang}, \citenamefont {Agarwal},\ and\ \citenamefont
  {Fleming}(2001)}]{Yang2001}%
  \BibitemOpen
  \bibfield  {author} {\bibinfo {author} {\bibfnamefont {M.}~\bibnamefont
  {Yang}}, \bibinfo {author} {\bibfnamefont {R.}~\bibnamefont {Agarwal}}, \
  and\ \bibinfo {author} {\bibfnamefont {G.~R.}\ \bibnamefont {Fleming}},\
  }\href {\doibase 10.1016/S1010-6030(01)00504-4} {\bibfield  {journal}
  {\bibinfo  {journal} {Journal of Photochemistry and Photobiology
  A-chemistry}\ }\textbf {\bibinfo {volume} {142}},\ \bibinfo {pages} {107}
  (\bibinfo {year} {2001})}\BibitemShut {NoStop}%
\bibitem [{\citenamefont {van Grondelle}\ and\ \citenamefont
  {Novoderezhkin}(2006)}]{Grondelle2006}%
  \BibitemOpen
  \bibfield  {author} {\bibinfo {author} {\bibfnamefont {R.}~\bibnamefont {van
  Grondelle}}\ and\ \bibinfo {author} {\bibfnamefont {V.~I.}\ \bibnamefont
  {Novoderezhkin}},\ }\href {\doibase 10.1039/b514032c} {\bibfield  {journal}
  {\bibinfo  {journal} {Physical Chemistry Chemical Physics}\ }\textbf
  {\bibinfo {volume} {8}},\ \bibinfo {pages} {793} (\bibinfo {year}
  {2006})}\BibitemShut {NoStop}%
\bibitem [{\citenamefont {Cogdell}, \citenamefont {Gall},\ and\ \citenamefont
  {Kohler}(2006)}]{Cogdell2006}%
  \BibitemOpen
  \bibfield  {author} {\bibinfo {author} {\bibfnamefont {R.~J.}\ \bibnamefont
  {Cogdell}}, \bibinfo {author} {\bibfnamefont {A.}~\bibnamefont {Gall}}, \
  and\ \bibinfo {author} {\bibfnamefont {J.}~\bibnamefont {Kohler}},\ }\href
  {\doibase 10.1017/S0033583506004434} {\bibfield  {journal} {\bibinfo
  {journal} {Q Rev Biophys}\ }\textbf {\bibinfo {volume} {39}},\ \bibinfo
  {pages} {227} (\bibinfo {year} {2006})}\BibitemShut {NoStop}%
\bibitem [{\citenamefont {Hu}\ \emph {et~al.}(2002)\citenamefont {Hu},
  \citenamefont {Ritz}, \citenamefont {Damjanovic}, \citenamefont
  {Autenrieth},\ and\ \citenamefont {Schulten}}]{Hu2002}%
  \BibitemOpen
  \bibfield  {author} {\bibinfo {author} {\bibfnamefont {X.~C.}\ \bibnamefont
  {Hu}}, \bibinfo {author} {\bibfnamefont {T.}~\bibnamefont {Ritz}}, \bibinfo
  {author} {\bibfnamefont {A.}~\bibnamefont {Damjanovic}}, \bibinfo {author}
  {\bibfnamefont {F.}~\bibnamefont {Autenrieth}}, \ and\ \bibinfo {author}
  {\bibfnamefont {K.}~\bibnamefont {Schulten}},\ }\href {\doibase
  10.1017/S0033583501003754} {\bibfield  {journal} {\bibinfo  {journal} {Q Rev
  Biophys}\ }\textbf {\bibinfo {volume} {35}},\ \bibinfo {pages} {1} (\bibinfo
  {year} {2002})}\BibitemShut {NoStop}%
\bibitem [{\citenamefont {Andrews}\ and\ \citenamefont
  {Demidov}(1999)}]{Andrew-book}%
  \BibitemOpen
  \bibfield  {author} {\bibinfo {author} {\bibfnamefont {D.~L.}\ \bibnamefont
  {Andrews}}\ and\ \bibinfo {author} {\bibfnamefont {A.~A.}\ \bibnamefont
  {Demidov}},\ }\href@noop {} {\emph {\bibinfo {title} {Resonance Energy
  Transfer}}}\ (\bibinfo  {publisher} {Wiley},\ \bibinfo {year}
  {1999})\BibitemShut {NoStop}%
\bibitem [{\citenamefont {Engel}\ \emph {et~al.}(2007)\citenamefont {Engel},
  \citenamefont {Calhoun}, \citenamefont {Read}, \citenamefont {Ahn},
  \citenamefont {Mancal}, \citenamefont {Cheng}, \citenamefont {Blankenship},\
  and\ \citenamefont {Fleming}}]{Engel-2007}%
  \BibitemOpen
  \bibfield  {author} {\bibinfo {author} {\bibfnamefont {G.~S.}\ \bibnamefont
  {Engel}}, \bibinfo {author} {\bibfnamefont {T.~R.}\ \bibnamefont {Calhoun}},
  \bibinfo {author} {\bibfnamefont {E.~L.}\ \bibnamefont {Read}}, \bibinfo
  {author} {\bibfnamefont {T.~K.}\ \bibnamefont {Ahn}}, \bibinfo {author}
  {\bibfnamefont {T.}~\bibnamefont {Mancal}}, \bibinfo {author} {\bibfnamefont
  {Y.~C.}\ \bibnamefont {Cheng}}, \bibinfo {author} {\bibfnamefont {R.~E.}\
  \bibnamefont {Blankenship}}, \ and\ \bibinfo {author} {\bibfnamefont {G.~R.}\
  \bibnamefont {Fleming}},\ }\href@noop {} {\bibfield  {journal} {\bibinfo
  {journal} {Nature}\ }\textbf {\bibinfo {volume} {446}},\ \bibinfo {pages}
  {782} (\bibinfo {year} {2007})}\BibitemShut {NoStop}%
\bibitem [{\citenamefont {Mohseni}\ \emph {et~al.}(2008)\citenamefont
  {Mohseni}, \citenamefont {Rebentrost}, \citenamefont {Lloyd},\ and\
  \citenamefont {Aspuru-Guzik}}]{Alan-2008}%
  \BibitemOpen
  \bibfield  {author} {\bibinfo {author} {\bibfnamefont {M.}~\bibnamefont
  {Mohseni}}, \bibinfo {author} {\bibfnamefont {P.}~\bibnamefont {Rebentrost}},
  \bibinfo {author} {\bibfnamefont {S.}~\bibnamefont {Lloyd}}, \ and\ \bibinfo
  {author} {\bibfnamefont {A.}~\bibnamefont {Aspuru-Guzik}},\ }\href@noop {}
  {\bibfield  {journal} {\bibinfo  {journal} {J Chem Phys}\ }\textbf {\bibinfo
  {volume} {129}},\ \bibinfo {pages} {174106} (\bibinfo {year}
  {2008})}\BibitemShut {NoStop}%
\bibitem [{\citenamefont {Rebentrost}\ \emph {et~al.}(2009)\citenamefont
  {Rebentrost}, \citenamefont {Mohseni}, \citenamefont {Kassal}, \citenamefont
  {Lloyd},\ and\ \citenamefont {Aspuru-Guzik}}]{Alan-2009}%
  \BibitemOpen
  \bibfield  {author} {\bibinfo {author} {\bibfnamefont {P.}~\bibnamefont
  {Rebentrost}}, \bibinfo {author} {\bibfnamefont {M.}~\bibnamefont {Mohseni}},
  \bibinfo {author} {\bibfnamefont {I.}~\bibnamefont {Kassal}}, \bibinfo
  {author} {\bibfnamefont {S.}~\bibnamefont {Lloyd}}, \ and\ \bibinfo {author}
  {\bibfnamefont {A.}~\bibnamefont {Aspuru-Guzik}},\ }\href@noop {} {\bibfield
  {journal} {\bibinfo  {journal} {New J Phys}\ }\textbf {\bibinfo {volume}
  {11}},\ \bibinfo {pages} {033003} (\bibinfo {year} {2009})}\BibitemShut
  {NoStop}%
\bibitem [{\citenamefont {Rebentrost}, \citenamefont {Mohseni},\ and\
  \citenamefont {Aspuru-Guzik}(2009)}]{Alan-JPC-2009}%
  \BibitemOpen
  \bibfield  {author} {\bibinfo {author} {\bibfnamefont {P.}~\bibnamefont
  {Rebentrost}}, \bibinfo {author} {\bibfnamefont {M.}~\bibnamefont {Mohseni}},
  \ and\ \bibinfo {author} {\bibfnamefont {A.}~\bibnamefont {Aspuru-Guzik}},\
  }\href@noop {} {\bibfield  {journal} {\bibinfo  {journal} {J Phys Chem B}\
  }\textbf {\bibinfo {volume} {113}},\ \bibinfo {pages} {9942} (\bibinfo {year}
  {2009})}\BibitemShut {NoStop}%
\bibitem [{\citenamefont {Rebentrost}\ and\ \citenamefont
  {Aspuru-Guzik}(2011)}]{Alan2011Com}%
  \BibitemOpen
  \bibfield  {author} {\bibinfo {author} {\bibfnamefont {P.}~\bibnamefont
  {Rebentrost}}\ and\ \bibinfo {author} {\bibfnamefont {A.}~\bibnamefont
  {Aspuru-Guzik}},\ }\href {\doibase 10.1063/1.3563617} {\bibfield  {journal}
  {\bibinfo  {journal} {J Chem Phys}\ }\textbf {\bibinfo {volume} {134}},\
  \bibinfo {pages} {101103} (\bibinfo {year} {2011})}\BibitemShut {NoStop}%
\bibitem [{\citenamefont {Ishizaki}\ and\ \citenamefont
  {Fleming}(2009{\natexlab{a}})}]{Fleming-JCP-2009}%
  \BibitemOpen
  \bibfield  {author} {\bibinfo {author} {\bibfnamefont {A.}~\bibnamefont
  {Ishizaki}}\ and\ \bibinfo {author} {\bibfnamefont {G.~R.}\ \bibnamefont
  {Fleming}},\ }\href@noop {} {\bibfield  {journal} {\bibinfo  {journal} {J
  Chem Phys}\ }\textbf {\bibinfo {volume} {130}},\ \bibinfo {pages} {234111}
  (\bibinfo {year} {2009}{\natexlab{a}})}\BibitemShut {NoStop}%
\bibitem [{\citenamefont {Ishizaki}\ and\ \citenamefont
  {Fleming}(2009{\natexlab{b}})}]{Fleming-PNAS-2009}%
  \BibitemOpen
  \bibfield  {author} {\bibinfo {author} {\bibfnamefont {A.}~\bibnamefont
  {Ishizaki}}\ and\ \bibinfo {author} {\bibfnamefont {G.~R.}\ \bibnamefont
  {Fleming}},\ }\href@noop {} {\bibfield  {journal} {\bibinfo  {journal} {Proc
  Natl Acad Sci U S A}\ }\textbf {\bibinfo {volume} {106}},\ \bibinfo {pages}
  {17255} (\bibinfo {year} {2009}{\natexlab{b}})}\BibitemShut {NoStop}%
\bibitem [{\citenamefont {Skochdopole}\ and\ \citenamefont
  {Mazziotti}(2011)}]{Mazziotti2011}%
  \BibitemOpen
  \bibfield  {author} {\bibinfo {author} {\bibfnamefont {N.}~\bibnamefont
  {Skochdopole}}\ and\ \bibinfo {author} {\bibfnamefont {D.~A.}\ \bibnamefont
  {Mazziotti}},\ }\href {\doibase 10.1021/jz201154t} {\bibfield  {journal}
  {\bibinfo  {journal} {J Phys Chem Lett}\ }\textbf {\bibinfo {volume} {2}},\
  \bibinfo {pages} {2989} (\bibinfo {year} {2011})},\ \Eprint
  {http://arxiv.org/abs/http://pubs.acs.org/doi/pdf/10.1021/jz201154t}
  {http://pubs.acs.org/doi/pdf/10.1021/jz201154t} \BibitemShut {NoStop}%
\bibitem [{\citenamefont {Sarovar}\ \emph {et~al.}(2010)\citenamefont
  {Sarovar}, \citenamefont {Ishizaki}, \citenamefont {Fleming},\ and\
  \citenamefont {Whaley}}]{Sarovar:2010p6945}%
  \BibitemOpen
  \bibfield  {author} {\bibinfo {author} {\bibfnamefont {M.}~\bibnamefont
  {Sarovar}}, \bibinfo {author} {\bibfnamefont {A.}~\bibnamefont {Ishizaki}},
  \bibinfo {author} {\bibfnamefont {G.~R.}\ \bibnamefont {Fleming}}, \ and\
  \bibinfo {author} {\bibfnamefont {K.~B.}\ \bibnamefont {Whaley}},\
  }\href@noop {} {\bibfield  {journal} {\bibinfo  {journal} {Nature Physics}\
  }\textbf {\bibinfo {volume} {6}},\ \bibinfo {pages} {462 } (\bibinfo {year}
  {2010})}\BibitemShut {NoStop}%
\bibitem [{\citenamefont {Shi}\ \emph {et~al.}(2009)\citenamefont {Shi},
  \citenamefont {Chen}, \citenamefont {Nan}, \citenamefont {Xu},\ and\
  \citenamefont {Yan}}]{Shi2009a}%
  \BibitemOpen
  \bibfield  {author} {\bibinfo {author} {\bibfnamefont {Q.}~\bibnamefont
  {Shi}}, \bibinfo {author} {\bibfnamefont {L.~P.}\ \bibnamefont {Chen}},
  \bibinfo {author} {\bibfnamefont {G.~J.}\ \bibnamefont {Nan}}, \bibinfo
  {author} {\bibfnamefont {R.~X.}\ \bibnamefont {Xu}}, \ and\ \bibinfo {author}
  {\bibfnamefont {Y.~J.}\ \bibnamefont {Yan}},\ }\href@noop {} {\bibfield
  {journal} {\bibinfo  {journal} {J Chem Phys}\ }\textbf {\bibinfo {volume}
  {130}},\ \bibinfo {pages} {084105} (\bibinfo {year} {2009})}\BibitemShut
  {NoStop}%
\bibitem [{\citenamefont {Zhu}\ \emph {et~al.}(2011{\natexlab{a}})\citenamefont
  {Zhu}, \citenamefont {Kais}, \citenamefont {Rebentrost},\ and\ \citenamefont
  {Aspuru-Guzik}}]{Ourpaper}%
  \BibitemOpen
  \bibfield  {author} {\bibinfo {author} {\bibfnamefont {J.}~\bibnamefont
  {Zhu}}, \bibinfo {author} {\bibfnamefont {S.}~\bibnamefont {Kais}}, \bibinfo
  {author} {\bibfnamefont {P.}~\bibnamefont {Rebentrost}}, \ and\ \bibinfo
  {author} {\bibfnamefont {A.}~\bibnamefont {Aspuru-Guzik}},\ }\href@noop {}
  {\bibfield  {journal} {\bibinfo  {journal} {J Phys Chem B}\ }\textbf
  {\bibinfo {volume} {115}},\ \bibinfo {pages} {1531} (\bibinfo {year}
  {2011}{\natexlab{a}})}\BibitemShut {NoStop}%
\bibitem [{\citenamefont {Zhu}\ \emph {et~al.}(2012)\citenamefont {Zhu},
  \citenamefont {Kais}, \citenamefont {Aspuru-Guzik}, \citenamefont
  {Rodriques}, \citenamefont {Brock},\ and\ \citenamefont {Love}}]{Ourpaper2}%
  \BibitemOpen
  \bibfield  {author} {\bibinfo {author} {\bibfnamefont {J.}~\bibnamefont
  {Zhu}}, \bibinfo {author} {\bibfnamefont {S.}~\bibnamefont {Kais}}, \bibinfo
  {author} {\bibfnamefont {A.}~\bibnamefont {Aspuru-Guzik}}, \bibinfo {author}
  {\bibfnamefont {S.}~\bibnamefont {Rodriques}}, \bibinfo {author}
  {\bibfnamefont {B.}~\bibnamefont {Brock}}, \ and\ \bibinfo {author}
  {\bibfnamefont {P.~J.}\ \bibnamefont {Love}},\ }\href@noop {} {\bibfield
  {journal} {\bibinfo  {journal} {arXiv:1202.4519v1}\ } (\bibinfo {year}
  {2012})},\ \Eprint {http://arxiv.org/abs/1202.4519} {1202.4519} \BibitemShut
  {NoStop}%
\bibitem [{\citenamefont {Kramer}\ \emph {et~al.}(2011)\citenamefont {Kramer},
  \citenamefont {Kreisbeck}, \citenamefont {Rodriguez},\ and\ \citenamefont
  {Hein}}]{kramerGPU}%
  \BibitemOpen
  \bibfield  {author} {\bibinfo {author} {\bibfnamefont {T.}~\bibnamefont
  {Kramer}}, \bibinfo {author} {\bibfnamefont {C.}~\bibnamefont {Kreisbeck}},
  \bibinfo {author} {\bibfnamefont {M.}~\bibnamefont {Rodriguez}}, \ and\
  \bibinfo {author} {\bibfnamefont {B.}~\bibnamefont {Hein}},\ }\href@noop {}
  {\  (\bibinfo {year} {2011})},\ \bibinfo {note} {american Physical Society
  March Meeting}\BibitemShut {NoStop}%
\bibitem [{\citenamefont {Berkelbach}, \citenamefont {Markland},\ and\
  \citenamefont {Reichman}(2011)}]{Reichman2011}%
  \BibitemOpen
  \bibfield  {author} {\bibinfo {author} {\bibfnamefont {T.~C.}\ \bibnamefont
  {Berkelbach}}, \bibinfo {author} {\bibfnamefont {T.~E.}\ \bibnamefont
  {Markland}}, \ and\ \bibinfo {author} {\bibfnamefont {D.~R.}\ \bibnamefont
  {Reichman}},\ }\href@noop {} {\bibfield  {journal} {\bibinfo  {journal}
  {arXiv: 1111.5026v1}\ } (\bibinfo {year} {2011})},\ \Eprint
  {http://arxiv.org/abs/1111.5026} {1111.5026} \BibitemShut {NoStop}%
\bibitem [{\citenamefont {Prior}\ \emph {et~al.}(2010)\citenamefont {Prior},
  \citenamefont {Chin}, \citenamefont {Huelga},\ and\ \citenamefont
  {Plenio}}]{PlenioDMRG2}%
  \BibitemOpen
  \bibfield  {author} {\bibinfo {author} {\bibfnamefont {J.}~\bibnamefont
  {Prior}}, \bibinfo {author} {\bibfnamefont {A.~W.}\ \bibnamefont {Chin}},
  \bibinfo {author} {\bibfnamefont {S.~F.}\ \bibnamefont {Huelga}}, \ and\
  \bibinfo {author} {\bibfnamefont {M.~B.}\ \bibnamefont {Plenio}},\ }\href
  {\doibase 10.1103/PhysRevLett.105.050404} {\bibfield  {journal} {\bibinfo
  {journal} {Phys Rev Lett}\ }\textbf {\bibinfo {volume} {105}},\ \bibinfo
  {pages} {050404} (\bibinfo {year} {2010})}\BibitemShut {NoStop}%
\bibitem [{\citenamefont {Huo}\ and\ \citenamefont {Coker}(2010)}]{coker2010}%
  \BibitemOpen
  \bibfield  {author} {\bibinfo {author} {\bibfnamefont {P.}~\bibnamefont
  {Huo}}\ and\ \bibinfo {author} {\bibfnamefont {D.~F.}\ \bibnamefont
  {Coker}},\ }\href {\doibase 10.1063/1.3498901} {\bibfield  {journal}
  {\bibinfo  {journal} {J Chem Phys}\ }\textbf {\bibinfo {volume} {133}},\
  \bibinfo {pages} {184108} (\bibinfo {year} {2010})}\BibitemShut {NoStop}%
\bibitem [{\citenamefont {Moix}\ \emph {et~al.}(2011)\citenamefont {Moix},
  \citenamefont {Wu}, \citenamefont {Huo}, \citenamefont {Coker},\ and\
  \citenamefont {Cao}}]{coker2011}%
  \BibitemOpen
  \bibfield  {author} {\bibinfo {author} {\bibfnamefont {J.}~\bibnamefont
  {Moix}}, \bibinfo {author} {\bibfnamefont {J.}~\bibnamefont {Wu}}, \bibinfo
  {author} {\bibfnamefont {P.}~\bibnamefont {Huo}}, \bibinfo {author}
  {\bibfnamefont {D.}~\bibnamefont {Coker}}, \ and\ \bibinfo {author}
  {\bibfnamefont {J.}~\bibnamefont {Cao}},\ }\href {\doibase 10.1021/jz201259v}
  {\bibfield  {journal} {\bibinfo  {journal} {J Phys Chem Lett}\ }\textbf
  {\bibinfo {volume} {2}},\ \bibinfo {pages} {3045} (\bibinfo {year}
  {2011})}\BibitemShut {NoStop}%
\bibitem [{\citenamefont {Mazziotti}(2011)}]{Mazziotti2011a}%
  \BibitemOpen
  \bibfield  {author} {\bibinfo {author} {\bibfnamefont {D.~A.}\ \bibnamefont
  {Mazziotti}},\ }\href@noop {} {\bibfield  {journal} {\bibinfo  {journal}
  {arXiv: 1112.5863v1}\ } (\bibinfo {year} {2011})},\ \Eprint
  {http://arxiv.org/abs/1112.5863} {1112.5863} \BibitemShut {NoStop}%
\bibitem [{\citenamefont {Nalbach}, \citenamefont {Braun},\ and\ \citenamefont
  {Thorwart}(2011)}]{Thorwart2011}%
  \BibitemOpen
  \bibfield  {author} {\bibinfo {author} {\bibfnamefont {P.}~\bibnamefont
  {Nalbach}}, \bibinfo {author} {\bibfnamefont {D.}~\bibnamefont {Braun}}, \
  and\ \bibinfo {author} {\bibfnamefont {M.}~\bibnamefont {Thorwart}},\ }\href
  {\doibase 10.1103/PhysRevE.84.041926} {\bibfield  {journal} {\bibinfo
  {journal} {Phys Rev E}\ }\textbf {\bibinfo {volume} {84}},\ \bibinfo {pages}
  {041926} (\bibinfo {year} {2011})}\BibitemShut {NoStop}%
\bibitem [{\citenamefont {Shabani}\ \emph {et~al.}(2011)\citenamefont
  {Shabani}, \citenamefont {Mohseni}, \citenamefont {Rabitz},\ and\
  \citenamefont {Lloyd}}]{Lloyd2011}%
  \BibitemOpen
  \bibfield  {author} {\bibinfo {author} {\bibfnamefont {A.}~\bibnamefont
  {Shabani}}, \bibinfo {author} {\bibfnamefont {M.}~\bibnamefont {Mohseni}},
  \bibinfo {author} {\bibfnamefont {H.}~\bibnamefont {Rabitz}}, \ and\ \bibinfo
  {author} {\bibfnamefont {S.}~\bibnamefont {Lloyd}},\ }\href@noop {}
  {\bibfield  {journal} {\bibinfo  {journal} {arXiv: 1103.3823v3}\ } (\bibinfo
  {year} {2011})},\ \Eprint {http://arxiv.org/abs/1103.3823} {1103.3823}
  \BibitemShut {NoStop}%
\bibitem [{\citenamefont {Mohseni}\ \emph {et~al.}(2011)\citenamefont
  {Mohseni}, \citenamefont {Shabani}, \citenamefont {Lloyd},\ and\
  \citenamefont {Rabitz}}]{Lloyd2011a}%
  \BibitemOpen
  \bibfield  {author} {\bibinfo {author} {\bibfnamefont {M.}~\bibnamefont
  {Mohseni}}, \bibinfo {author} {\bibfnamefont {A.}~\bibnamefont {Shabani}},
  \bibinfo {author} {\bibfnamefont {S.}~\bibnamefont {Lloyd}}, \ and\ \bibinfo
  {author} {\bibfnamefont {H.}~\bibnamefont {Rabitz}},\ }\href@noop {}
  {\bibfield  {journal} {\bibinfo  {journal} {arXiv: 1104.4812v1}\ } (\bibinfo
  {year} {2011})},\ \Eprint {http://arxiv.org/abs/1104.4812} {1104.4812}
  \BibitemShut {NoStop}%
\bibitem [{\citenamefont {Lloyd}\ \emph {et~al.}(2011)\citenamefont {Lloyd},
  \citenamefont {Mohseni}, \citenamefont {Shabani},\ and\ \citenamefont
  {Rabitz}}]{Lloyd2011b}%
  \BibitemOpen
  \bibfield  {author} {\bibinfo {author} {\bibfnamefont {S.}~\bibnamefont
  {Lloyd}}, \bibinfo {author} {\bibfnamefont {M.}~\bibnamefont {Mohseni}},
  \bibinfo {author} {\bibfnamefont {A.}~\bibnamefont {Shabani}}, \ and\
  \bibinfo {author} {\bibfnamefont {H.}~\bibnamefont {Rabitz}},\ }\href@noop {}
  {\bibfield  {journal} {\bibinfo  {journal} {arXiv:1111.4982v1}\ } (\bibinfo
  {year} {2011})},\ \Eprint {http://arxiv.org/abs/1111.4982} {1111.4982}
  \BibitemShut {NoStop}%
\bibitem [{\citenamefont {Kim}\ and\ \citenamefont {Cao}(2010)}]{cao2010}%
  \BibitemOpen
  \bibfield  {author} {\bibinfo {author} {\bibfnamefont {J.~H.}\ \bibnamefont
  {Kim}}\ and\ \bibinfo {author} {\bibfnamefont {J.~S.}\ \bibnamefont {Cao}},\
  }\href {\doibase 10.1021/jp106838k} {\bibfield  {journal} {\bibinfo
  {journal} {J Phys Chem B}\ }\textbf {\bibinfo {volume} {114}},\ \bibinfo
  {pages} {16189} (\bibinfo {year} {2010})}\BibitemShut {NoStop}%
\bibitem [{\citenamefont {Wu}\ \emph {et~al.}(2011)\citenamefont {Wu},
  \citenamefont {Liu}, \citenamefont {Ma}, \citenamefont {Silbey},\ and\
  \citenamefont {Cao}}]{cao2011}%
  \BibitemOpen
  \bibfield  {author} {\bibinfo {author} {\bibfnamefont {J.}~\bibnamefont
  {Wu}}, \bibinfo {author} {\bibfnamefont {F.}~\bibnamefont {Liu}}, \bibinfo
  {author} {\bibfnamefont {J.}~\bibnamefont {Ma}}, \bibinfo {author}
  {\bibfnamefont {R.~J.}\ \bibnamefont {Silbey}}, \ and\ \bibinfo {author}
  {\bibfnamefont {J.}~\bibnamefont {Cao}},\ }\href@noop {} {\bibfield
  {journal} {\bibinfo  {journal} {arXiv:1109.5769v1}\ } (\bibinfo {year}
  {2011})},\ \Eprint {http://arxiv.org/abs/1109.5769} {1109.5769} \BibitemShut
  {NoStop}%
\bibitem [{\citenamefont {Wu}\ \emph {et~al.}(2010)\citenamefont {Wu},
  \citenamefont {Liu}, \citenamefont {Shen}, \citenamefont {Cao},\ and\
  \citenamefont {Silbey}}]{Silbey2010}%
  \BibitemOpen
  \bibfield  {author} {\bibinfo {author} {\bibfnamefont {J.~L.}\ \bibnamefont
  {Wu}}, \bibinfo {author} {\bibfnamefont {F.}~\bibnamefont {Liu}}, \bibinfo
  {author} {\bibfnamefont {Y.}~\bibnamefont {Shen}}, \bibinfo {author}
  {\bibfnamefont {J.~S.}\ \bibnamefont {Cao}}, \ and\ \bibinfo {author}
  {\bibfnamefont {R.~J.}\ \bibnamefont {Silbey}},\ }\href {\doibase
  10.1088/1367-2630/12/10/105012} {\bibfield  {journal} {\bibinfo  {journal}
  {New J Phys}\ }\textbf {\bibinfo {volume} {12}},\ \bibinfo {pages} {105012}
  (\bibinfo {year} {2010})}\BibitemShut {NoStop}%
\bibitem [{\citenamefont {Vlaming}\ and\ \citenamefont
  {Silbey}(2011)}]{Silbey2011}%
  \BibitemOpen
  \bibfield  {author} {\bibinfo {author} {\bibfnamefont {S.~M.}\ \bibnamefont
  {Vlaming}}\ and\ \bibinfo {author} {\bibfnamefont {R.~J.}\ \bibnamefont
  {Silbey}},\ }\href@noop {} {\bibfield  {journal} {\bibinfo  {journal} {arXiv:
  1111.3627v1}\ } (\bibinfo {year} {2011})},\ \Eprint
  {http://arxiv.org/abs/1111.3627} {1111.3627} \BibitemShut {NoStop}%
\bibitem [{\citenamefont {Renaud}, \citenamefont {Ratner},\ and\ \citenamefont
  {Mujica}(2011)}]{Ratner2011}%
  \BibitemOpen
  \bibfield  {author} {\bibinfo {author} {\bibfnamefont {N.}~\bibnamefont
  {Renaud}}, \bibinfo {author} {\bibfnamefont {M.~A.}\ \bibnamefont {Ratner}},
  \ and\ \bibinfo {author} {\bibfnamefont {V.}~\bibnamefont {Mujica}},\ }\href
  {\doibase 10.1063/1.3624376} {\bibfield  {journal} {\bibinfo  {journal} {J
  Chem Phys}\ }\textbf {\bibinfo {volume} {135}},\ \bibinfo {pages} {075102}
  (\bibinfo {year} {2011})}\BibitemShut {NoStop}%
\bibitem [{\citenamefont {Abramavicius}\ and\ \citenamefont
  {Mukamel}(2010)}]{Mukamel2010a}%
  \BibitemOpen
  \bibfield  {author} {\bibinfo {author} {\bibfnamefont {D.}~\bibnamefont
  {Abramavicius}}\ and\ \bibinfo {author} {\bibfnamefont {S.}~\bibnamefont
  {Mukamel}},\ }\href {\doibase 10.1063/1.3458824} {\bibfield  {journal}
  {\bibinfo  {journal} {J Chem Phys}\ }\textbf {\bibinfo {volume} {133}},\
  \bibinfo {pages} {064510} (\bibinfo {year} {2010})}\BibitemShut {NoStop}%
\bibitem [{\citenamefont {Scholes}(2010)}]{Scholes-2010}%
  \BibitemOpen
  \bibfield  {author} {\bibinfo {author} {\bibfnamefont {G.~D.}\ \bibnamefont
  {Scholes}},\ }\href@noop {} {\bibfield  {journal} {\bibinfo  {journal} {J
  Phys Chem Lett}\ }\textbf {\bibinfo {volume} {1}},\ \bibinfo {pages} {2}
  (\bibinfo {year} {2010})}\BibitemShut {NoStop}%
\bibitem [{\citenamefont {Chachisvilis}\ \emph {et~al.}(1997)\citenamefont
  {Chachisvilis}, \citenamefont {Kuhn}, \citenamefont {Pullerits},\ and\
  \citenamefont {Sundstrom}}]{Chachisvilis1997}%
  \BibitemOpen
  \bibfield  {author} {\bibinfo {author} {\bibfnamefont {M.}~\bibnamefont
  {Chachisvilis}}, \bibinfo {author} {\bibfnamefont {O.}~\bibnamefont {Kuhn}},
  \bibinfo {author} {\bibfnamefont {T.}~\bibnamefont {Pullerits}}, \ and\
  \bibinfo {author} {\bibfnamefont {V.}~\bibnamefont {Sundstrom}},\ }\href
  {\doibase 10.1021/jp963360a} {\bibfield  {journal} {\bibinfo  {journal} {J
  Phys Chem B}\ }\textbf {\bibinfo {volume} {101}},\ \bibinfo {pages} {7275}
  (\bibinfo {year} {1997})}\BibitemShut {NoStop}%
\bibitem [{\citenamefont {Harel}\ and\ \citenamefont
  {Engel}(2012)}]{Harel2012}%
  \BibitemOpen
  \bibfield  {author} {\bibinfo {author} {\bibfnamefont {E.}~\bibnamefont
  {Harel}}\ and\ \bibinfo {author} {\bibfnamefont {G.~S.}\ \bibnamefont
  {Engel}},\ }\href {\doibase 10.1073/pnas.1110312109} {\bibfield  {journal}
  {\bibinfo  {journal} {Proc Natl Acad Sci U S A}\ }\textbf {\bibinfo {volume}
  {109}},\ \bibinfo {pages} {706} (\bibinfo {year} {2012})}\BibitemShut
  {NoStop}%
\bibitem [{\citenamefont {Strumpfer}\ and\ \citenamefont
  {Schulten}(2009)}]{Strumpfer2009a}%
  \BibitemOpen
  \bibfield  {author} {\bibinfo {author} {\bibfnamefont {J.}~\bibnamefont
  {Strumpfer}}\ and\ \bibinfo {author} {\bibfnamefont {K.}~\bibnamefont
  {Schulten}},\ }\href {\doibase 10.1063/1.3271348} {\bibfield  {journal}
  {\bibinfo  {journal} {J Chem Phys}\ }\textbf {\bibinfo {volume} {131}},\
  \bibinfo {pages} {225101} (\bibinfo {year} {2009})}\BibitemShut {NoStop}%
\bibitem [{\citenamefont {Strumpfer}\ and\ \citenamefont
  {Schulten}(2011)}]{Strumpfer2011}%
  \BibitemOpen
  \bibfield  {author} {\bibinfo {author} {\bibfnamefont {J.}~\bibnamefont
  {Strumpfer}}\ and\ \bibinfo {author} {\bibfnamefont {K.}~\bibnamefont
  {Schulten}},\ }\href {\doibase 10.1063/1.3557042} {\bibfield  {journal}
  {\bibinfo  {journal} {J Chem Phys}\ }\textbf {\bibinfo {volume} {134}},\
  \bibinfo {pages} {095102} (\bibinfo {year} {2011})}\BibitemShut {NoStop}%
\bibitem [{\citenamefont {Papiz}\ \emph {et~al.}(2003)\citenamefont {Papiz},
  \citenamefont {Prince}, \citenamefont {Howard}, \citenamefont {Cogdell},\
  and\ \citenamefont {Isaacs}}]{Papiz2003}%
  \BibitemOpen
  \bibfield  {author} {\bibinfo {author} {\bibfnamefont {M.}~\bibnamefont
  {Papiz}}, \bibinfo {author} {\bibfnamefont {S.}~\bibnamefont {Prince}},
  \bibinfo {author} {\bibfnamefont {T.}~\bibnamefont {Howard}}, \bibinfo
  {author} {\bibfnamefont {R.}~\bibnamefont {Cogdell}}, \ and\ \bibinfo
  {author} {\bibfnamefont {N.}~\bibnamefont {Isaacs}},\ }\href {\doibase
  10.1016/S0022-2836(03)00024-X} {\bibfield  {journal} {\bibinfo  {journal} {J
  Mol Biol}\ }\textbf {\bibinfo {volume} {326}},\ \bibinfo {pages} {1523}
  (\bibinfo {year} {2003})}\BibitemShut {NoStop}%
\bibitem [{\citenamefont {Koepke}\ \emph {et~al.}(1996)\citenamefont {Koepke},
  \citenamefont {Hu}, \citenamefont {Muenke}, \citenamefont {Schulten},\ and\
  \citenamefont {Michel}}]{Koepke1996}%
  \BibitemOpen
  \bibfield  {author} {\bibinfo {author} {\bibfnamefont {J.}~\bibnamefont
  {Koepke}}, \bibinfo {author} {\bibfnamefont {X.}~\bibnamefont {Hu}}, \bibinfo
  {author} {\bibfnamefont {C.}~\bibnamefont {Muenke}}, \bibinfo {author}
  {\bibfnamefont {K.}~\bibnamefont {Schulten}}, \ and\ \bibinfo {author}
  {\bibfnamefont {H.}~\bibnamefont {Michel}},\ }\href {\doibase
  10.1016/S0969-2126(96)00063-9} {\bibfield  {journal} {\bibinfo  {journal}
  {Structure}\ }\textbf {\bibinfo {volume} {4}},\ \bibinfo {pages} {581}
  (\bibinfo {year} {1996})}\BibitemShut {NoStop}%
\bibitem [{\citenamefont {MCDERMOTT}\ \emph {et~al.}(1995)\citenamefont
  {MCDERMOTT}, \citenamefont {PRINCE}, \citenamefont {FREER}, \citenamefont
  {HAWTHORNTHWAITELAWLESS}, \citenamefont {PAPIZ}, \citenamefont {COGDELL},\
  and\ \citenamefont {ISAACS}}]{MCDERMOTT1995}%
  \BibitemOpen
  \bibfield  {author} {\bibinfo {author} {\bibfnamefont {G.}~\bibnamefont
  {MCDERMOTT}}, \bibinfo {author} {\bibfnamefont {S.}~\bibnamefont {PRINCE}},
  \bibinfo {author} {\bibfnamefont {A.}~\bibnamefont {FREER}}, \bibinfo
  {author} {\bibfnamefont {A.}~\bibnamefont {HAWTHORNTHWAITELAWLESS}}, \bibinfo
  {author} {\bibfnamefont {M.}~\bibnamefont {PAPIZ}}, \bibinfo {author}
  {\bibfnamefont {R.}~\bibnamefont {COGDELL}}, \ and\ \bibinfo {author}
  {\bibfnamefont {N.}~\bibnamefont {ISAACS}},\ }\href {\doibase
  10.1038/374517a0} {\bibfield  {journal} {\bibinfo  {journal} {Nature}\
  }\textbf {\bibinfo {volume} {374}},\ \bibinfo {pages} {517} (\bibinfo {year}
  {1995})}\BibitemShut {NoStop}%
\bibitem [{\citenamefont {Sener}\ and\ \citenamefont
  {Schulten}(2008)}]{PurpleBac}%
  \BibitemOpen
  \bibfield  {author} {\bibinfo {author} {\bibfnamefont {M.}~\bibnamefont
  {Sener}}\ and\ \bibinfo {author} {\bibfnamefont {K.}~\bibnamefont
  {Schulten}},\ }\href@noop {} {\emph {\bibinfo {title} {The Purple
  Phototrophic Bacteria, of Advances in Photosynthesis and Respiration}}},\
  edited by\ \bibinfo {editor} {\bibfnamefont {M.~T.}\ \bibnamefont
  {C.N.~Hunter}, \bibfnamefont {F.~Daldal}}\ and\ \bibinfo {editor}
  {\bibfnamefont {J.}~\bibnamefont {Beatty}},\ Vol.~\bibinfo {volume} {28}\
  (\bibinfo  {publisher} {Springer, New York},\ \bibinfo {year} {2008})\ pp.\
  \bibinfo {pages} {275--294}\BibitemShut {NoStop}%
\bibitem [{\citenamefont {Zhang}\ \emph {et~al.}(1998)\citenamefont {Zhang},
  \citenamefont {Meier}, \citenamefont {Chernyak},\ and\ \citenamefont
  {Mukamel}}]{redfield1}%
  \BibitemOpen
  \bibfield  {author} {\bibinfo {author} {\bibfnamefont {W.~M.}\ \bibnamefont
  {Zhang}}, \bibinfo {author} {\bibfnamefont {T.}~\bibnamefont {Meier}},
  \bibinfo {author} {\bibfnamefont {V.}~\bibnamefont {Chernyak}}, \ and\
  \bibinfo {author} {\bibfnamefont {S.}~\bibnamefont {Mukamel}},\ }\href
  {\doibase 10.1063/1.476212} {\bibfield  {journal} {\bibinfo  {journal}
  {Journal of Chemical Physics}\ }\textbf {\bibinfo {volume} {108}},\ \bibinfo
  {pages} {7763} (\bibinfo {year} {1998})}\BibitemShut {NoStop}%
\bibitem [{\citenamefont {Yang}\ and\ \citenamefont
  {Fleming}(2002)}]{redfield2}%
  \BibitemOpen
  \bibfield  {author} {\bibinfo {author} {\bibfnamefont {M.}~\bibnamefont
  {Yang}}\ and\ \bibinfo {author} {\bibfnamefont {G.~R.}\ \bibnamefont
  {Fleming}},\ }\href {\doibase 10.1016/S0301-0104(02)00603-1} {\bibfield
  {journal} {\bibinfo  {journal} {Chemical Physics}\ }\textbf {\bibinfo
  {volume} {282}},\ \bibinfo {pages} {PII S0301} (\bibinfo {year}
  {2002})}\BibitemShut {NoStop}%
\bibitem [{\citenamefont {B.~W. Van Der~Meer}\ and\ \citenamefont
  {Chen}(1994)}]{ForsterBook1}%
  \BibitemOpen
  \bibfield  {author} {\bibinfo {author} {\bibfnamefont {G.~C.~I.}\
  \bibnamefont {B.~W. Van Der~Meer}}\ and\ \bibinfo {author} {\bibfnamefont
  {S.-Y.~S.}\ \bibnamefont {Chen}},\ }\href@noop {} {\emph {\bibinfo {title}
  {Resonance Energy Transfer: Theory and Data}}}\ (\bibinfo  {publisher} {John
  Wiley \& Sons},\ \bibinfo {year} {1994})\BibitemShut {NoStop}%
\bibitem [{\citenamefont {Agranovich}\ and\ \citenamefont
  {Hochstrasser}(1982)}]{ForsterBook2}%
  \BibitemOpen
  \bibfield  {author} {\bibinfo {author} {\bibfnamefont {V.}~\bibnamefont
  {Agranovich}}\ and\ \bibinfo {author} {\bibfnamefont {M.}~\bibnamefont
  {Hochstrasser}},\ }\href@noop {} {\emph {\bibinfo {title} {Electronic
  Excitation Energy Transfer in Condensed Matter}}}\ (\bibinfo  {publisher}
  {Elsevier Science Ltd},\ \bibinfo {year} {1982})\BibitemShut {NoStop}%
\bibitem [{\citenamefont {Tanimura}(2006)}]{Tanimura}%
  \BibitemOpen
  \bibfield  {author} {\bibinfo {author} {\bibfnamefont {Y.}~\bibnamefont
  {Tanimura}},\ }\href@noop {} {\bibfield  {journal} {\bibinfo  {journal} {J
  Phys Soc Jpn}\ }\textbf {\bibinfo {volume} {75}},\ \bibinfo {pages} {082001}
  (\bibinfo {year} {2006})}\BibitemShut {NoStop}%
\bibitem [{\citenamefont {Tanimura}\ and\ \citenamefont
  {Kubo}(1989)}]{Tanimura-Kubo-1989}%
  \BibitemOpen
  \bibfield  {author} {\bibinfo {author} {\bibfnamefont {Y.}~\bibnamefont
  {Tanimura}}\ and\ \bibinfo {author} {\bibfnamefont {R.}~\bibnamefont
  {Kubo}},\ }\href@noop {} {\bibfield  {journal} {\bibinfo  {journal} {J Phys
  Soc Jpn}\ }\textbf {\bibinfo {volume} {58}},\ \bibinfo {pages} {101}
  (\bibinfo {year} {1989})}\BibitemShut {NoStop}%
\bibitem [{\citenamefont {Ishizaki}\ and\ \citenamefont
  {Tanimura}(2005)}]{Tanimura2}%
  \BibitemOpen
  \bibfield  {author} {\bibinfo {author} {\bibfnamefont {A.}~\bibnamefont
  {Ishizaki}}\ and\ \bibinfo {author} {\bibfnamefont {Y.}~\bibnamefont
  {Tanimura}},\ }\href@noop {} {\bibfield  {journal} {\bibinfo  {journal} {J
  Phys Soc Jpn}\ }\textbf {\bibinfo {volume} {74}},\ \bibinfo {pages} {3131}
  (\bibinfo {year} {2005})}\BibitemShut {NoStop}%
\bibitem [{\citenamefont {Renger}, \citenamefont {May},\ and\ \citenamefont
  {Kuhn}(2001)}]{Renger2001}%
  \BibitemOpen
  \bibfield  {author} {\bibinfo {author} {\bibfnamefont {T.}~\bibnamefont
  {Renger}}, \bibinfo {author} {\bibfnamefont {V.}~\bibnamefont {May}}, \ and\
  \bibinfo {author} {\bibfnamefont {O.}~\bibnamefont {Kuhn}},\ }\href@noop {}
  {\bibfield  {journal} {\bibinfo  {journal} {Physics Reports-review Section of
  Physics Letters}\ }\textbf {\bibinfo {volume} {343}},\ \bibinfo {pages} {138}
  (\bibinfo {year} {2001})}\BibitemShut {NoStop}%
\bibitem [{\citenamefont {Hu}\ \emph {et~al.}(1997)\citenamefont {Hu},
  \citenamefont {Ritz}, \citenamefont {Damjanovic},\ and\ \citenamefont
  {Schulten}}]{Hu1997}%
  \BibitemOpen
  \bibfield  {author} {\bibinfo {author} {\bibfnamefont {X.~C.}\ \bibnamefont
  {Hu}}, \bibinfo {author} {\bibfnamefont {T.}~\bibnamefont {Ritz}}, \bibinfo
  {author} {\bibfnamefont {A.}~\bibnamefont {Damjanovic}}, \ and\ \bibinfo
  {author} {\bibfnamefont {K.}~\bibnamefont {Schulten}},\ }\href {\doibase
  10.1021/jp963777g} {\bibfield  {journal} {\bibinfo  {journal} {J Phys Chem
  B}\ }\textbf {\bibinfo {volume} {101}},\ \bibinfo {pages} {3854} (\bibinfo
  {year} {1997})}\BibitemShut {NoStop}%
\bibitem [{\citenamefont {Sener}\ \emph {et~al.}(2007)\citenamefont {Sener},
  \citenamefont {Olsen}, \citenamefont {Hunter},\ and\ \citenamefont
  {Schulten}}]{Sener2007}%
  \BibitemOpen
  \bibfield  {author} {\bibinfo {author} {\bibfnamefont {M.~K.}\ \bibnamefont
  {Sener}}, \bibinfo {author} {\bibfnamefont {J.~D.}\ \bibnamefont {Olsen}},
  \bibinfo {author} {\bibfnamefont {C.~N.}\ \bibnamefont {Hunter}}, \ and\
  \bibinfo {author} {\bibfnamefont {K.}~\bibnamefont {Schulten}},\ }\href
  {\doibase 10.1073/pnas.0706861104} {\bibfield  {journal} {\bibinfo  {journal}
  {Proc Natl Acad Sci U S A}\ }\textbf {\bibinfo {volume} {104}},\ \bibinfo
  {pages} {15723} (\bibinfo {year} {2007})}\BibitemShut {NoStop}%
\bibitem [{\citenamefont {Tretiak}\ \emph {et~al.}(2000)\citenamefont
  {Tretiak}, \citenamefont {Middleton}, \citenamefont {Chernyak},\ and\
  \citenamefont {Mukamel}}]{Tretiak2000}%
  \BibitemOpen
  \bibfield  {author} {\bibinfo {author} {\bibfnamefont {S.}~\bibnamefont
  {Tretiak}}, \bibinfo {author} {\bibfnamefont {C.}~\bibnamefont {Middleton}},
  \bibinfo {author} {\bibfnamefont {V.}~\bibnamefont {Chernyak}}, \ and\
  \bibinfo {author} {\bibfnamefont {S.}~\bibnamefont {Mukamel}},\ }\href
  {\doibase 10.1021/jp001585m} {\bibfield  {journal} {\bibinfo  {journal} {J
  Phys Chem B}\ }\textbf {\bibinfo {volume} {104}},\ \bibinfo {pages} {9540}
  (\bibinfo {year} {2000})}\BibitemShut {NoStop}%
\bibitem [{\citenamefont {Zhu}\ \emph {et~al.}(2011{\natexlab{b}})\citenamefont
  {Zhu}, \citenamefont {Xu}, \citenamefont {Zhang}, \citenamefont {Hu},\ and\
  \citenamefont {Yan}}]{xu1}%
  \BibitemOpen
  \bibfield  {author} {\bibinfo {author} {\bibfnamefont {K.-B.}\ \bibnamefont
  {Zhu}}, \bibinfo {author} {\bibfnamefont {R.-X.}\ \bibnamefont {Xu}},
  \bibinfo {author} {\bibfnamefont {H.~Y.}\ \bibnamefont {Zhang}}, \bibinfo
  {author} {\bibfnamefont {J.}~\bibnamefont {Hu}}, \ and\ \bibinfo {author}
  {\bibfnamefont {Y.~J.}\ \bibnamefont {Yan}},\ }\href {\doibase
  10.1021/jp2002244} {\bibfield  {journal} {\bibinfo  {journal} {J Phys Chem
  B}\ }\textbf {\bibinfo {volume} {115}},\ \bibinfo {pages} {5678} (\bibinfo
  {year} {2011}{\natexlab{b}})}\BibitemShut {NoStop}%
\bibitem [{\citenamefont {Xu}\ \emph {et~al.}(2011)\citenamefont {Xu},
  \citenamefont {Xu}, \citenamefont {Abramavicius}, \citenamefont {Zhang},\
  and\ \citenamefont {Yan}}]{xu2}%
  \BibitemOpen
  \bibfield  {author} {\bibinfo {author} {\bibfnamefont {J.}~\bibnamefont
  {Xu}}, \bibinfo {author} {\bibfnamefont {R.-x.}\ \bibnamefont {Xu}}, \bibinfo
  {author} {\bibfnamefont {D.}~\bibnamefont {Abramavicius}}, \bibinfo {author}
  {\bibfnamefont {H.-d.}\ \bibnamefont {Zhang}}, \ and\ \bibinfo {author}
  {\bibfnamefont {Y.-j.}\ \bibnamefont {Yan}},\ }\href {\doibase
  10.1088/1674-0068/24/05/497-506} {\bibfield  {journal} {\bibinfo  {journal}
  {Chin J Chem Phys}\ }\textbf {\bibinfo {volume} {24}},\ \bibinfo {pages}
  {497} (\bibinfo {year} {2011})}\BibitemShut {NoStop}%
\bibitem [{\citenamefont {Freiberg}\ \emph {et~al.}(2009)\citenamefont
  {Freiberg}, \citenamefont {Ratsep}, \citenamefont {Timpmann},\ and\
  \citenamefont {Trinkunas}}]{Freiberg2009}%
  \BibitemOpen
  \bibfield  {author} {\bibinfo {author} {\bibfnamefont {A.}~\bibnamefont
  {Freiberg}}, \bibinfo {author} {\bibfnamefont {M.}~\bibnamefont {Ratsep}},
  \bibinfo {author} {\bibfnamefont {K.}~\bibnamefont {Timpmann}}, \ and\
  \bibinfo {author} {\bibfnamefont {G.}~\bibnamefont {Trinkunas}},\ }\href
  {\doibase 10.1016/j.chemphys.2008.10.043} {\bibfield  {journal} {\bibinfo
  {journal} {Chem Phys}\ }\textbf {\bibinfo {volume} {357}},\ \bibinfo {pages}
  {102} (\bibinfo {year} {2009})}\BibitemShut {NoStop}%
\bibitem [{\citenamefont {Zerlauskiene}\ \emph {et~al.}(2008)\citenamefont
  {Zerlauskiene}, \citenamefont {Trinkunas}, \citenamefont {Gall},
  \citenamefont {Robert}, \citenamefont {Urboniene},\ and\ \citenamefont
  {Valkunas}}]{Zerlauskiene2008}%
  \BibitemOpen
  \bibfield  {author} {\bibinfo {author} {\bibfnamefont {O.}~\bibnamefont
  {Zerlauskiene}}, \bibinfo {author} {\bibfnamefont {G.}~\bibnamefont
  {Trinkunas}}, \bibinfo {author} {\bibfnamefont {A.}~\bibnamefont {Gall}},
  \bibinfo {author} {\bibfnamefont {B.}~\bibnamefont {Robert}}, \bibinfo
  {author} {\bibfnamefont {V.}~\bibnamefont {Urboniene}}, \ and\ \bibinfo
  {author} {\bibfnamefont {L.}~\bibnamefont {Valkunas}},\ }\href {\doibase
  10.1021/jp803439w} {\bibfield  {journal} {\bibinfo  {journal} {J Phys Chem
  B}\ }\textbf {\bibinfo {volume} {112}},\ \bibinfo {pages} {15883} (\bibinfo
  {year} {2008})}\BibitemShut {NoStop}%
\bibitem [{\citenamefont {Tian}\ \emph {et~al.}(2010)\citenamefont {Tian},
  \citenamefont {Ding}, \citenamefont {Xu},\ and\ \citenamefont {Yan}}]{xu3}%
  \BibitemOpen
  \bibfield  {author} {\bibinfo {author} {\bibfnamefont {B.-L.}\ \bibnamefont
  {Tian}}, \bibinfo {author} {\bibfnamefont {J.-J.}\ \bibnamefont {Ding}},
  \bibinfo {author} {\bibfnamefont {R.-X.}\ \bibnamefont {Xu}}, \ and\ \bibinfo
  {author} {\bibfnamefont {Y.}~\bibnamefont {Yan}},\ }\href {\doibase
  10.1063/1.3491270} {\bibfield  {journal} {\bibinfo  {journal} {J Chem Phys}\
  }\textbf {\bibinfo {volume} {133}},\ \bibinfo {pages} {114112} (\bibinfo
  {year} {2010})}\BibitemShut {NoStop}%
\bibitem [{\citenamefont {Ding}\ \emph {et~al.}(2011)\citenamefont {Ding},
  \citenamefont {Xu}, \citenamefont {Hu}, \citenamefont {Xu},\ and\
  \citenamefont {Yan}}]{xu4}%
  \BibitemOpen
  \bibfield  {author} {\bibinfo {author} {\bibfnamefont {J.-J.}\ \bibnamefont
  {Ding}}, \bibinfo {author} {\bibfnamefont {J.}~\bibnamefont {Xu}}, \bibinfo
  {author} {\bibfnamefont {J.}~\bibnamefont {Hu}}, \bibinfo {author}
  {\bibfnamefont {R.-X.}\ \bibnamefont {Xu}}, \ and\ \bibinfo {author}
  {\bibfnamefont {Y.}~\bibnamefont {Yan}},\ }\href@noop {} {\bibfield
  {journal} {\bibinfo  {journal} {J Chem Phys}\ }\textbf {\bibinfo {volume}
  {135}},\ \bibinfo {pages} {164107} (\bibinfo {year} {2011})}\BibitemShut
  {NoStop}%
\end{thebibliography}
\end{document}